\RequirePackage{fix-cm}

\documentclass[draft]{elsarticle}
\usepackage{amssymb}
\usepackage[ruled,vlined]{algorithm2e}
\newtheorem{Definition}{\bf Definition}[section]

\newtheorem{Proposition}{\bf Proposition}[section]

\newtheorem{Theorem}{\bf Theorem}[section]

\newtheorem{Lemma}{\bf Lemma}[section]

\def\done{$\hfill\Box$}

\newcommand{\bform}{\begin{displaymath}}
\newcommand{\eform}{\end{displaymath}}
\newcommand{\beqn}[1]{\begin{equation}\label{#1}}
\newcommand{\eeqn}{\end{equation}\smallskip}
\newcommand{\beqna}[1]{\begin{eqnarray}\label{#1}}
\newcommand{\eeqna}{\end{eqnarray}\smallskip}
\newcommand{\beqnan}{\begin{eqnarray*}}
\newcommand{\eeqnan}{\end{eqnarray*}\smallskip}

\usepackage[english]{babel}

\begin{document}

\begin{frontmatter}

\title{An ${\cal O}(n\sqrt{m})$ algorithm for the weighted stable set problem in \{claw, net\}-free graphs with $\alpha(G) \ge 4$}

\author[salento]{Paolo Nobili}
\author[sapienza]{Antonio Sassano\corref{cor1}}
\ead{sassano@dis.uniroma1.it}
\cortext[cor1]{Corresponding author}
\address[salento]{Dipartimento di Ingegneria dell'Innovazione, Universit\`a del Salento, Lecce, Italy}
\address[sapienza]{Dipartimento di Informatica e Sistemistica ``Antonio Ruberti'',  Universit\`a di Roma ``La Sapienza'', Roma, Italy}

\begin{abstract}
In this paper we show that a connected \{claw, net\}-free graph $G(V, E)$ with $\alpha(G) \ge 4$ is the union of a \emph{strongly bisimplicial} clique $Q$ and at most two \emph{clique-strips}. A clique is strongly bisimplicial if its neighborhood is partitioned into two cliques which are mutually non-adjacent and a clique-strip is a sequence of cliques $\{H_0, \dots, H_p\}$ with the property that $H_i$ is adjacent only to $H_{i-1}$ and $H_{i+1}$. By exploiting such a structure we show how to solve the \emph{Maximum Weight Stable Set Problem} in such a graph in time ${\cal O}(|V|\sqrt{|E|})$, improving the previous complexity bound of ${\cal O}(|V||E|)$.
\end{abstract}


\end{frontmatter}

\section{Introduction}
%=====================

\noindent
The \emph{Maximum Weight Stable Set Problem (MWSSP)} in a graph $G(V, E)$ with node-weight function $w: V \rightarrow \Re$ asks for a subset $S^*$ of pairwise non-adjacent \emph{nodes} in $V$ having maximum weight $\sum_{v \in S^*} w(v) = \alpha_w (G)$. For each subset $W$ of $V$ we denote by $\alpha_w (W)$ the maximum weight of a stable set in $W$. If $w$ is the vector of all $1$'s we omit the reference to $w$ and write $\alpha (G)$ and $\alpha (W)$.

\bigskip\noindent
For each graph $G(V, E)$ we denote by $V(F)$ the set of end-nodes of the edges in $F \subseteq E$, by $E(W)$ the set of edges with end-nodes in $W \subseteq V$ and by $N(W)$ (\emph{neighborhood} of $W$) the set of nodes in $V \setminus W$ adjacent to some node in $W$. If $W = \{w\}$ we simply write $N(w)$. We denote by $N[W]$ and $N[w]$ (\emph{closed neighborhood}) the sets $N(W) \cup W$ and $N(w) \cup \{w\}$ and by $\delta(W)$ the set of edges having exactly one end-node in $W$; if $\delta(W) = \emptyset$ and $W$ is minimal with this property we say that $W$ is (or induces) a connected component of $G$. We denote by $G - F$ the subgraph of $G$ obtained by removing from $G$ the edges in $F \subseteq E$. A \emph{clique} is a complete subgraph of $G$ induced by some set of nodes $K \subseteq V$. With a little abuse of notation we also regard the set $K$ as a clique and, for any edge $uv \in E$, both $uv$ and $\{u, v\}$ are said to be a clique. A node $w$ such that $N(w)$ is a clique is said to be \emph{simplicial}. By extension, a clique $K$ such that $N(K)$ is a clique is also said to be \emph{simplicial}. A \emph{claw} is a graph with four nodes $w, x, y, z$ with $w$ adjacent to $x, y, z$ and $x, y, z$ mutually non-adjacent. To highlight its structure, it is denoted as $(w: x, y, z)$. A $P_k$ is a (chordless) path induced by $k$ nodes and will be denoted as $(u_1, \dots, u_k)$. A subset $T \in V$ is \emph{null} (\emph{universal}) to a subset $W \subseteq V \setminus T$ if and only if $N(T) \cap W = \emptyset$ ($N(T) \cap W = W$). Two nodes $u, v \in V$ are said to be \emph{twins} if $N(u) \setminus \{v\} = N(v) \setminus \{u\}$. We can always remove a twin from $V$ without affecting the value of the optimal solution of MWSSP. In fact, if $uv \in E$ we can remove the twin with minimum weight, while if $uv \notin E$ we can remove $u$ and replace $w(v)$ by $w(u) + w(v)$. The complexity of finding all the twins is ${\cal O}(|V|+|E|)$ (\cite{McConnellS94}, \cite{CournierH94}) and hence we assume throughout the paper that our graphs have no twins. A \emph{net} $(x, y, z: x', y', z')$ is a graph induced by a triangle $T = \{x, y, z\}$ and three mutually non-adjacent nodes $\{x', y', z'\}$ with $N(x') \cap T = \{x\}$, $N(y') \cap T = \{y\}$ and $N(z') \cap T = \{z\}$. A \emph{square} is a $4$-hole $(v_1, v_2, v_3, v_4)$ with $v_1 v_3, v_2 v_4 \notin E$ called \emph{diagonals}.

\medskip\noindent
The family of \{claw, net\}-free graphs has been widely studied in the literature (\cite{PullShep}, \cite{Brandstadt03}, \cite{OrioloetAl10}) since such graphs constitute an important subclass of claw-free graphs. In particular, in \cite{PullShep} Pulleyblank and Shepherd described both a ${\cal O}(|V|^4)$ algorithm for the maximum weight stable set problem in distance claw-free graphs (a class containing \{claw, net\}-free graphs) and the structure of a polyhedron whose projection gives the stable set polyhedron $STAB(G)$. In \cite{FaenzaOS14} Faenza, Oriolo and Stauffer reduced the complexity of MWSSP in \{claw, net\}-free graphs to ${\cal O}(|V||E|)$, which constitutes a bottleneck for the complexity of their algorithm for the MWSSP in claw-free graphs. In this paper we give a new structural characterization of \{claw, net\}-free graphs with stability number not smaller than four which allows us to define a ${\cal O}(|V|\sqrt{|E|})$ time algorithm for the MWSSP in such graphs. This result, together with the ${\cal O}(|E|\log{|V|})$ time algorithm for the MWSSP in claw-free graphs with stability number at most three described in \cite{NobiliSassano15a}, provides a ${\cal O}(\sqrt{|E|}(|V| + \sqrt{|E|}\log{|V|})$ time algorithm for the MWSSP in \{claw, net\}-free graphs which improves the result of \cite{FaenzaOS14}.

\medskip\noindent
We say that a node $v \in V$ is \emph{regular} if its neighborhood can be partitioned into two cliques. A maximal clique $Q$ is \emph{reducible} if $\alpha(N(Q)) \leq 2$. If $Q$ is a maximal clique, two non-adjacent nodes $u, v \in N(Q)$ are said to be \emph{$Q$-distant} if $N(u) \cap N(v) \cap Q = \emptyset$ and \emph{$Q$-close} otherwise ($N(u) \cap N(v) \cap Q \ne \emptyset$). A maximal clique $Q$ is \emph{normal} if it has three independent neighbors that are mutually $Q$-distant. In \cite{LovaszPlummer} Lov\'asz and Plummer proved the following useful properties of a maximal clique in a claw-free graph.

\begin{Proposition}\label{PropertiesNormalClique}
Let $G(V, E)$ be a claw-free graph. If $Q$ is a maximal clique in $G$ then:
\begin{description}
   \item[\emph{(i)}] if $u$ and $v$ are $Q$-close nodes then $Q \subseteq N(u) \cup N(v)$;
   \item[\emph{(ii)}] if $u, v, w$ are mutually non-adjacent nodes in $N(Q)$ and two of them are $Q$-distant then any two of them are $Q$-distant and hence $Q$ is normal.
\end{description}
\done
\end{Proposition}
\medskip\noindent

\begin{Theorem}\label{NotReducibleIsNormal}
Let $G(V, E)$ be a claw-free graph and $u$ a regular node in $V$ whose closed neighborhood is covered by two maximal cliques $Q$ and $\overline Q$. If $Q$ ($\overline Q$) is not reducible then it is normal.
\end{Theorem}

\noindent
\emph{Proof}. Suppose, by contradiction, that $Q$ is not normal and $\alpha(N(Q)) \ge 3$. Let $v_1, v_2, v_3$ be three mutually non-adjacent nodes in $N(Q)$. If $u$ is not adjacent to $v_1, v_2, v_3$, then by \emph{(i)} of Proposition~\ref{PropertiesNormalClique} we have that $v_1, v_2, v_3$ are mutually $Q$-distant and hence $Q$ is normal, a contradiction. Consequently, without loss of generality, we can assume that $u$ is adjacent to $v_1$ and so $v_1$ belongs to $\overline Q$. The nodes $v_2$ and $v_3$ do not belong to $Q \cup \overline Q$ and hence are not adjacent to $u$. It follows, again by \emph{(i)} of Proposition~\ref{PropertiesNormalClique}, that $v_2$ and $v_3$ are distant with respect to $Q$. But then, by \emph{(ii)} of Proposition~\ref{PropertiesNormalClique}, $Q$ is a normal clique, a contradiction.
\done

\section{Wings and similarity classes}
%=====================================

\smallskip\noindent
%
%%%% Risposta a commento minore 2 di referee 1
%
Let $G$ be a claw-free graph and let $S$ be a stable set of $G(V, E)$. Any node $s \in S$ is said to be \emph{stable}; any node $v \in V \setminus S$ satisfies $|N(v) \cap S| \le 2$ and is called \emph{superfree} if $|N(v) \cap S| = 0$, \emph{free} if $|N(v) \cap S| = 1$ and \emph{bound} if $|N(v) \cap S| = 2$. For each free node $u$ we denote by $S(u)$ the unique node in $S$ adjacent to $u$. Observe that, by claw-freeness, a bound node $b$ cannot be adjacent to a node $u \in V \setminus S$ unless $b$ and $u$ have a common neighbor in $S$.

\smallskip\noindent
We denote by $F(T)$ the set of free nodes with respect to $S$ which are adjacent to some node in $T \subseteq S$; to simplify our notation we will always write $F(s)$ instead of $F(\{s\})$.

\smallskip\noindent
A \emph{bound-wing} defined by $\{s, t\} \subseteq S$ is the set $W^B(s, t) = \{u \in V \setminus S: N(u) \cap S = \{s, t\}\}$ if non-empty. A \emph{free-wing} defined by the ordered pair $(s, t)$ ($s, t \in S$) is the set $W^F(s, t) = \{u \in F(s): N(u) \cap F(t) \ne \emptyset\}$ if non-empty. Observe that, by claw-freeness, any bound node is contained in a single bound-wing. On the contrary, a free node can belong to several free-wings. Moreover, while $W^B(s, t) \equiv W^B(t, s)$, we have $W^F(s, t) \ne W^F(t, s)$. By slightly generalizing the definition due to Minty \cite{Minty}, we call \emph{wing} defined by $(s, t)$ ($s, t \in S$) the set $W(s, t) = W^B(s, t) \cup W^F(s, t) \cup W^F(t, s)$ if non-empty. Observe that $W(s, t) = W(t, s)$. The nodes $s$ and $t$ are said to be the \emph{extrema} of the wing $W(s, t)$.

\medskip\noindent
We say that two nodes $u$ and $v$ in $V \setminus S$ are \emph{similar} ($u \sim v$) if $N(u) \cap S = N(v) \cap S$ and \emph{dissimilar}($u \not\sim v$) otherwise. Clearly, similarity induces an equivalence relation on $V \setminus S$ and a partition in \emph{similarity classes}. Similarity classes can be \emph{bound}, \emph{free} or \emph{superfree} in that they are entirely composed by nodes that are bound, free or superfree with respect to $S$. Bound similarity classes are precisely the bound-wings defined by pairs of nodes of $S$, while each free similarity class contains the (free) nodes adjacent to the same node of $S$. Let $V_F$ be the set of nodes that are free with respect to $S$ and let $G_F(V_F, E_F)$ be the graph with edge-set $E_F = \{uv \in E: u, v \in V_F, u \not\sim v\}$ (\emph{free dissimilarity graph}).

\begin{Definition}\label{FreeComponent}
Let $G(V, E)$ be a claw-free graph, $S$ a maximal stable set in $G$ and $G_F$ the free dissimilarity graph of $G$ with respect to $S$. A connected component of $G_F$ inducing a maximal clique in $G$ is said to be a \emph{free component} of $G$ with respect to $S$.
\done
\end{Definition}

\begin{Theorem}\label{Similar}
Let $G(V, E)$ be a claw-free graph and $S$ a maximal stable set of $G$. Then a connected component of $G_F$ intersecting three or more free similarity classes induces a maximal clique in $G$ and hence is a free component.
\end{Theorem}

\noindent
\emph{Proof}. We first claim that the nodes of any chordless path $P$ in $G_F$, the free dissimilarity graph of $G$ with respect to $S$, connecting two dissimilar nodes $u, v$ belong only to the similarity classes of $u$ and $v$. In fact, two consecutive nodes of $P$ necessarily belong to different classes. If a node in a third class existed in $P$ we would necessarily have three consecutive nodes $x, y, z$ of $P$ in three different classes. But then $(y: S(y), x, z)$ would be a claw in $G$, a contradiction. Suppose now that a connected component $X$ of $G_F$, intersecting three or more similarity classes, is not a clique in $G$ and let $u, z \in X$ be two nonadjacent nodes in $G$. Suppose first that $S(u)$ and $S(z)$ are two distinct nodes of $S$ and, consequently, that $u \not\sim z$. Let $v \in X$ be a node with $S(v) \notin \{S(u), S(z)\}$, it exists since we assumed that $X$ intersects more than two similarity classes. Let $P_{uv}$ and $P_{vz}$ be chordless paths connecting $u$ to $v$ and, respectively, $v$ to $z$ in $G_F$. By the above claim, $P_{uv}$ contains only nodes in the similarity classes of $u$ and $v$, while $P_{vz}$ contains only nodes in the similarity classes of $v$ and $z$. Let $W_{uz}$ be the walk connecting $u$ to $z$ obtained by chaining $P_{uv}$ and $P_{vz}$ and let $P_{uz}$ be any chordless path connecting $u$ to $z$ whose nodes belong to $W_{uz}$. Since $uz \notin E$, $P_{uz}$ contains at least one node in the similarity class of $v$ and hence contains nodes in three different similarity classes, contradicting the hypothesis that $P_{uz}$ is chordless. It follows that $u$ and $z$ belong to the same similarity class. More generally, any two dissimilar nodes in $X$ are adjacent. Let $v \in X$ be a node with $S(v) \ne S(u) \equiv S(z)$. It follows that $uv, vz \in E$ and hence $(v: S(v), u, z)$ is a claw, a contradiction. It follows that $X$ is a clique in $G$. To prove that it is also maximal, assume by contradiction that there exists some node $u \in N(X)$ universal to $X$. The node $u$ is not free for, otherwise, it would belong to $X$ and is not stable since $X$ intersects more than one similarity class. It follows that $u$ is bound and adjacent to two nodes $s, t \in S$. Moreover, there exists some node $z \in N(X) \cap S$ with $z \ne s, t$. But then, for each node $x \in X \cap N(z)$, $(u: s, t, x)$ is a claw in $G$, a contradiction.
\done

\medskip\noindent
The previous theorem applies to claw-free graphs in general. However, in the special case of \{claw, net\}-free graphs, we have the following.

\begin{Proposition}\label{PropertiesWings}
Let $G(V, E)$ be a \{claw, net\}-free graph and $S$ a maximal stable set. Then any node in $V \setminus S$ is contained in a single wing.
\end{Proposition}

\noindent
\emph{Proof}. Let $u$ be a node in $V \setminus S$. Since $S$ is maximal, $u$ is either bound or free. In the first case $N(u) \cap S = \{s, t\}$ and $u$ belongs to $W(s, t)$. In the second case suppose, by contradiction, that $u$ belongs to the wings $W(S(u), t_1)$ and $W(S(u), t_2)$ with $t_1 \ne t_2$. Let $u_1 \in F(t_1)$ and $u_2 \in F(t_2)$ be nodes adjacent to $u$. If $u_1$ is not adjacent to $u_2$ then $(u: S(u), u_1, u_2)$ is a claw in $G$, while if $u_1$ is adjacent to $u_2$ then $(u, u_1, u_2: S(u), t_1, t_2)$ is a net in $G$. In both cases we have a contradiction.
\done

\section{Canonical Stable Sets}
%=================================

\noindent
Here we introduce a special class of maximal stable sets in $G$ which will be instrumental in this paper. In what follows a free node $x$ with $N[x] \supsetneq N[S(x)]$ will be called a \emph{dominating free} node, while the node $S(x)$ will be said to be \emph{dominated} by $x$.

\begin{Definition}\label{CanonicalStableSet}
Let $G(V, E)$ be a connected claw-free graph. A maximal stable set $S$ of $G$ is said to be \emph{canonical} if and only if $G$ does not contain:

\begin{description}
   \item[\emph{(i)}] a $P_3$ $(x, s, y)$, where $x$ and $y$ are free and $s$ is stable with respect to $S$ (augmenting $P_3$ with respect to $S$);
   \item[\emph{(ii)}] a dominating free node.
\done
\end{description}
\end{Definition}

\medskip\noindent
A stable set satisfying conditions \emph{(i)} and \emph{(ii)} of Definition~\ref{CanonicalStableSet} can be easily obtained from a maximal stable set $S_0$ of a connected claw-free graph $G$ by repeatedly applying the following two operations:

\begin{description}
   \item[\emph{(a)}] \emph{augmentation} along an augmenting $P_3$ $(x, s, y)$ with respect to $S$ ($S := S \cup \{x, y\} \setminus \{s\}$);
   \item[\emph{(b)}] \emph{alternation} along a $P_2$ $(x, S(x))$ (an edge), where $x$ is a dominating free node of $S(x)$ with $|N[x]| \ge |N[x']|$ for each free node $x'$ dominating $S(x)$ ($S := S \cup \{x\} \setminus \{S(x)\}$).
\end{description}

\begin{Theorem}\label{CanonicalConstruction}
Let $G(V, E)$ be a claw-free graph and $S_0$ a maximal stable set of $G$. Then a canonical stable set can be obtained from $S_0$ by repeatedly applying operations \emph{(a)} and \emph{(b)} in time ${\cal O}(|E|)$.
\end{Theorem}

\noindent
\emph{Proof}. Let $S$ be any maximal stable set of $G$. We first prove the following claims.

\medskip\noindent
\emph{Claim (i). Let $T$ be a stable set obtained from $S$ by applying operation \emph{(a)} along a $P_3$ $(\bar x, s, \bar y)$; then the set of free nodes with respect to $T$ is a proper subset of the set of free nodes with respect to $S$ and every $P_3$ augmenting with respect to $T$ is also augmenting with respect to $S$, in particular it does not contain $\bar x$ or $\bar y$.}

\smallskip\noindent
\emph{Proof.} Suppose first that there exists some node $x$ which is free with respect to $T$ but is not free with respect to $S$. Since $s$ is bound with respect to $T$ and $S$ is maximal, we have that $x \ne s$ is bound with respect to $S$. The node $x$ is adjacent to $s$ (otherwise it would be bound also with respect to $T$) and to some other stable node $\bar s \in S$. Moreover, since $x$ is free with respect to $T$ and is adjacent to $\bar s \in T$, it is non-adjacent to $\bar x$ and to $\bar y$. But then $(s: \bar x, \bar y, x)$ is a claw in $G$, a contradiction. Suppose now that there exists an augmenting $P_3$ $(x, t, y)$ with respect to $T$ that is not augmenting with respect to $S$. Since $x$ and $y$ are free nodes also with respect to $S$, we have that $t$ is in $T \setminus S$, $x$ and $y$ are non-adjacent to any node in $S \cap T$ and hence are both adjacent to $s$. Hence, without loss of generality, we can assume $t \equiv \bar x$ and so $\bar y$ is non-adjacent to both $x$ and $y$. But then $(s: x, y, \bar y)$ is a claw in $G$, a contradiction.

\noindent
\emph{End of Claim (i).}

\medskip\noindent
\emph{Claim (ii). Let $x$ be a dominating free node having maximum degree among the free nodes dominating $S(x)$. If $T$ is a stable set obtained from $S$ by applying operation \emph{(b)} along $(x, S(x))$ and there is no augmenting $P_3$ with respect to $S$, then there is no augmenting $P_3$ with respect to $T$. Moreover, every dominating free node with respect to $T$ is also dominating with respect to $S$ and does not dominate $x$.}

\smallskip\noindent
\emph{Proof.} Observe that the alternation along the path $(x, S(x))$ does not create new free nodes, since every node adjacent to $S(x)$ is also adjacent to $x$. Assume that an augmenting $P_3$ $(y, t, z)$ exists with respect to $T$, while no augmenting $P_3$ exists with respect to $S$. We have that $y$ and $z$ are also free with respect to $S$. If $t$ is in $S$, then $(y, t, z)$ is also augmenting with respect to $S$. Otherwise, $t \equiv x$ and $y, z$ are necessarily in $N(S(x))$, so we have the augmenting path $(y, S(x), z)$ with respect to $S$. In both cases we contradict the assumption that no augmenting $P_3$ exists with respect to $S$. Now let $t$ be a dominating free node with respect to $T$. Since $t$ is also a free node with respect to $S$, if $T(t) \ne x$ then $t$ is dominating free also with respect to $S$. Assume, conversely, that $T(t) = x$. Since $t$ dominates $x$ with respect to $T$, it satisfies $N(t) \supset N(x) \setminus \{t\}$. In particular, $t$ is adjacent to $S(x)$ and all of its neighbors and so it dominates $S(x)$  with respect to $S$. But this violates the assumption that $x$ has maximum degree among the free nodes dominating $S(x)$. The claim follows.

\noindent
\emph{End of Claim (ii).}

\medskip\noindent
Let $S_0 = \{s_1, s_2, \dots, s_q\}$ be a maximal stable set of $G(V, E)$ and let $F_0$ be the set of free nodes with respect to $S_0$. We now prove that a stable set $Z_0$ such that no augmenting $P_3$ exists in $G$ with respect to it can be obtained from $S_0$ by repeatedly applying operation \emph{(a)} to a current stable set $S$ (initialized as $S := S_0$) in overall time ${\cal O}(|E|)$. Let $F$ be the set of free nodes with respect to $S$. At any stage of the procedure we examine a node $s_i \in S_0$. Let $G_i(V_i, E_i)$ be the subgraph of $G$ induced by $N[s_i]$. Observe that, by \cite{KKM00}, $|V_i| = {\cal O}(\sqrt{|E_i|})$. We scan the set $V_i \cap F$ looking for a pair of non-adjacent nodes. This can be done in time ${\cal O}(|E_i|)$. If we find an augmenting $P_3$ $(x_i, s_i, y_i)$, we update the stable set $S$ by performing operation \emph{(a)} ($S := S \setminus \{s_i\} \cup \{x_i, y_i\}$). Moreover, the set $F$ is updated by removing $x_i$, $y_i$, any node adjacent to $x_i$ or $y_i$ and not to $s_i$ (every such node is necessarily free with respect to the former stable set and becomes bound after operation \emph{(a)} is performed) and any node adjacent to both $x_i$ and $y_i$ (every such node must be free with respect to the former stable set, is adjacent to $s_i$ and becomes bound after operation \emph{(a)} is performed). It follows that $F$ can be updated in time ${\cal O}(|\delta(x_i) \cup \delta(y_i)|)$. Observe that, for each node $u \in F \setminus V_i$, $S(u)$ is not changed by operation \emph{(a)}. In addition, Claim~\emph{(i)} ensures that no new augmenting $P_3$ is produced by the operation. In particular, no $P_3$ augmenting with respect to $S$ exists with $x_i$ ($y_i$) as stable node. This implies that we have only to check the nodes in $S_0$ as stable nodes in augmenting $P_3$ and we can avoid to update $S(u)$ for any free node $u \in N(s_i)$. It follows that the overall complexity of the procedure is ${\cal O}(\sum_{i = 1}^{q}(|E_i| + |\delta(x_i) \cup \delta(y_i)|) = {\cal O}(|E|)$.

\medskip\noindent
Let $\{s_1, s_2, \dots, s_r\}$ be the nodes in $Z_0$. We now prove that a canonical stable set can be obtained from $Z_0$ by repeatedly applying operation \emph{(b)} to a current stable set $S$ (initialized as $S := Z_0$) in overall time ${\cal O}(|E|)$. Let $F$ be the set of free nodes with respect to $S$. At any stage of the procedure we examine a node $s_i \in Z_0$. Let $G_i(V_i, E_i)$ be the subgraph of $G$ induced by $N[s_i]$ and let $F_i = V_i \cap F$. As observed above, $|V_i| = {\cal O}(\sqrt{|E_i|})$. To produce the set of free nodes dominating $s_i$, we remove from $F_i$ the nodes that have some non-adjacent node in $V_i$. To this purpose we scan all the pairs of non-adjacent nodes in $V_i$. This can be done in time ${\cal O}(|E_i|)$. Now, if the resulting $F_i$ is non-empty, we let $x_i$ be any node in $F_i$ having maximum degree and perform operation \emph{(b)} ($S := S \setminus \{s_i\} \cup \{x_i\}$). Moreover, we update the set $F$ by removing $x_i$ and any node adjacent to $x_i$ and not to $s_i$ (every such node is necessarily free with respect to the former stable set and becomes bound after operation \emph{(b)} is performed). It follows that $F$ can be updated in time ${\cal O}(|\delta(x_i)|)$. By Claim~\emph{(ii)} the new stable set does not induce an augmenting $P_3$ or new dominating free nodes and no free node in $N(x_i)$ can be dominating. This implies that only the nodes in $Z_0$ have to be checked for domination. It follows that the overall complexity of the procedure is ${\cal O}(\sum_{i = 1}^{r}(|E_i| + |\delta(x_i)|) = {\cal O}(|E|)$. Hence, in time ${\cal O}(|E|)$ we obtain a canonical stable set and the theorem follows.
\done

\section{The structure of \{claw, net\}-free graphs}
%===================================================

\noindent
Let $S_3^-$ be the graph with nodes $\{a, b, c, d, e, f\}$ and edges $\{ad, ae, be, bf, cd, cf,$ $de, df\}$. In their extensive study of the structure of \{claw, net\}-free graphs, Brandst{\"a}dt and Dragan have proved the following crucial result.

\begin{Lemma}\label{Brandstadt}\cite{Brandstadt03}
Let $G(V, E)$ be a \{claw, net\}-free graph and let $H \subset V$ be a set of nodes inducing an $S_3^-$. Then every node in $V \setminus H$ is adjacent to two nodes in $H$.
\end{Lemma}

\medskip\noindent
In \cite{LovaszPlummer} Lov\'asz and Plummer proved that every irregular node $a$ of a claw-free graph satisfies $\alpha (\{a\} \cup N(a) \cup N^2 (a)) \le 3$, where $N^2(a) = N(N(a)) \setminus \{a\}$. We use this fact to prove the following lemma.

\begin{Lemma}\label{RegularityLemma}
Let $G(V, E)$ be a connected \{claw, net\}-free graph with $\alpha(G) \ge 4$. Then each node in $G$ is regular.
\end{Lemma}

\noindent
\emph{Proof}. Assume, by contradiction, that $G$ contains an irregular node $a$ and let $H = \{v_1, \dots, v_{2k+1}\}$ be an odd anti-hole in $N(a)$, where $\{v_i, \dots, v_{i+k-1}\}$ is a clique and $\{v_i, v_{i+k}\}$, $\{v_i, v_{i+k+1}\}$ are stable sets, for each $i \in \{1, \dots, 2k+1\}$ (sums taken modulo $2k+1$). We claim that $N^2(H) = N(N(H)) \setminus H$ is empty. Otherwise, let $x$ be a node in $N^2(H)$ and let $y \in N(H)$ be a node adjacent to $x$. By claw-freeness, $N(y) \cap H$ is a clique $Q$. Without loss of generality, we can assume $y v_1 \in E$ and $Q \subseteq \{v_1, \dots, v_k\}$. Let $v_j$ be the node in $Q$ with largest index. If $j < k$ then $(v_1: y, v_{2k+1}, v_k)$ is a claw in $G$, a contradiction. Hence we have $j = k$. But then $(y, v_1, v_k: x, v_{2k+1}, v_{k+1})$ is a net in $G$, again a contradiction. It follows that $V = H \cup N(H) = \{a\} \cup N(a) \cup N^2 (a)$, contradicting the fact that $\alpha (\{a\} \cup N(a) \cup N^2 (a)) \le 3$.
\done

\begin{Theorem}\label{3WingedNormal}
Let $G(V, E)$ be a \{claw, net\}-free graph with $\alpha(G) \ge 4$ and $S$ a canonical stable set in $G$. Then each node $s \in S$ defines at most two wings.
\end{Theorem}

\noindent
\emph{Proof}. Assume, by contradiction, that there exists a node $s \in S$ defining at least three wings. By Lemma~\ref{RegularityLemma}, the node $s$ is regular; let $C_s, \overline C_s$ be any pair of maximal cliques covering $N[s]$. Let $W(s, t_i)$ ($i = 1, 2, 3$) be three wings intersecting $N(s)$. Without loss of generality, we can assume that two of them, say $W(s, t_1)$ and $W(s, t_2)$, intersect $C_s$.

\medskip\noindent
\emph{Claim (i). For each pair $W(s, t_i)$, $W(s, t_j)$ ($i \ne j$) intersecting $C_s$ and nodes $x, y \in N(C_s) \setminus \overline C_s$ with $x \in \{t_i\} \cup W(s, t_i)$ and $y \in \{t_j\} \cup W(s, t_j)$ we have that $x$ and $y$ are $C_s$-distant.}

%
%%%%%%%%%%%%%%%%%%%%%%%%%%%%%%%%%%%%%%%%%%%%%%%%%%%
%
%   Modifica 24/9/2015 - Punto 1 Reviewer 1
%
\smallskip\noindent
\emph{Proof.} Since $x, y \in N(C_s) \setminus \overline C_s$ we have $xs, ys \notin E$ and hence $x \in F(t_i) \cup \{t_i\}$ and $y \in F(t_j) \cup \{t_j\}$. Assume, by contradiction, that $x$ and $y$ have a common neighbor $z \in C_s$. Since $z \ne s$ we have $xy \in E$ (otherwise $(z: x, y, s)$ would be a claw in $G$). It follows that $x \ne t_i$ and $y \ne t_j$. But then $(x, y, z: t_i, t_j, s)$ is a net in $G$, a contradiction. The claim follows.

\noindent
\emph{End of Claim (i).}

\medskip\noindent
Since $W(s, t_1)$ and $W(s, t_2)$ intersect $C_s$, there exist two nodes $x, y \in N(C_s) \setminus \overline C_s$ with $x \in \{t_1\} \cup W(s, t_1)$ and $y \in \{t_2\} \cup W(s, t_2)$. By Claim~\emph{(i)}, $x$ and $y$ are $C_s$-distant. If there exists a node $z \in \overline C_s \setminus C_s$ which is non-adjacent to both $x$ and $y$ then, by \emph{(ii)} of Proposition~\ref{PropertiesNormalClique}, $C_s$ is normal, a contradiction. If, conversely, every node in $\overline C_s \setminus C_s$ is either adjacent to $x$ or to $y$ then $\overline C_s \setminus C_s \subseteq W(s, t_1) \cup W(s, t_2)$ and, by Proposition~\ref{PropertiesWings}, $W(s, t_3)$ does not intersect $\overline C_s  \setminus C_s$. But then $W(s, t_3)$ intersects $C_s$ and there exists a node $z \in N(C_s) \setminus \overline C_s$ with $z \in \{t_3\} \cup W(s, t_3)$. By Claim~\emph{(i)}, $x$, $y$ and $z$ are mutually $C_s$-distant and hence $C_s$ is normal, again a contradiction. The theorem follows.
\done
%
%%%%%%%%%%%%%%%%%%%%%%%%%%%%%%%%%%%%%%%%%%%%%%%%%%%
%

\begin{Definition}\label{StronglyBisimplicial}
A maximal clique $Q$ in a graph $G(V, E)$ is said to be \emph{bisimplicial} if $N(Q)$ is partitioned into two cliques $K_1$, $K_2$. The clique $Q$ is said to be \emph{strongly bisimplicial} if $K_1$ is null to $K_2$ and \emph{dominating} if
%
%%%%%%%%%%%%%%%%%%%%%%%%%%%%%%%%%%%%%%%%%%%%%%%%%%%
%
%   Modifica 3/9/2015
%
%   there exists an edge $uv \in E$ with the property that $u \in K_1$, $v \in K_2$ and $N(\{u, v\}) \supseteq V \setminus Q$.
each edge $uv \in E$ with $u \in K_1$ and $v \in K_2$ satisfies $N[\{u, v\}] \supseteq V \setminus Q$.
%
%%%%%%%%%%%%%%%%%%%%%%%%%%%%%%%%%%%%%%%%%%%%%%%%%%%
%
\done
\end{Definition}

\medskip\noindent
Observe that, in particular, a maximal simplicial clique is strongly bisimplicial. Oriolo, Pietropaoli and Stauffer introduced the following useful notion.

\begin{Definition}\cite{OrioloetAl08}\label{SemiHomogeneous}
Let $G(V, E)$ be a graph. A pair of cliques $(X, Y)$ in $G$ is \emph{semi-homogeneous} if for all $u \in V \setminus (X \cup Y)$, $u$ is either universal to $X$ or universal to $Y$ or null to $X \cup Y$.
\done
\end{Definition}

\medskip\noindent
In what follows we will use a weakening of the above definition.

\begin{Definition}\label{C4SemiHomogeneous}
Let $G(V, E)$ be a graph. A pair of disjoint cliques $(A, B)$ in $G$ is \emph{square-semi-homogeneous} if for any induced square $C$ contained in $A \cup B$ the sets $C \cap A$ and $C \cap B$ define a semi-homogeneous pair of cliques in $G$.
\done
\end{Definition}

\medskip\noindent
Examples of pairs of cliques which are square-semi-homogeneous but not semi-homogeneous can be easily constructed. In particular consider two triangles $K_1 = \{a_1, b_1, c_1\}$ and $K_2 = \{a_2, b_2, c_2\}$ with $a_1$ adjacent to $b_2$ and $c_2$, $b_1$ adjacent to $b_2$ and $c_1$ adjacent to $c_2$. If there exists a node $v$ null to $K_2$ and adjacent to $b_1$ and $c_1$ in $K_1$ then the pair $(K_1, K_2)$ is square-semi-homogeneous but not semi-homogeneous. Two useful properties of square-semi-homogeneous pairs of cliques are described by the following theorem.

\begin{Theorem}\label{C4Properties}
%
%%%%%%%%%%%%%%%%%%%%%%%%%%%%%%%%%%%%%%%%%%%%%%%%%%%
%
%   Modifica 7/9/2015
%
Let $G(V, E)$ be a connected \{claw, net\}-free graph with $\alpha(G) \ge 4$ and let $(A, B)$ be a
square-semi-homogeneous pair of cliques in $G$ with the property that each node in $A$ is non-null to $B$ and viceversa. Let $a_{max} \in A$ be a node which maximizes $|N(a) \cap B|$ for $a \in A$.
%
%   and $V \setminus (H \cup K)$ is either null to $H$ or to $K$
%
%%%%%%%%%%%%%%%%%%%%%%%%%%%%%%%%%%%%%%%%%%%%%%%%%%%
%
Then the following properties hold:
\begin{description}
%
%%%%%%%%%%%%%%%%%%%%%%%%%%%%%%%%%%%%%%%%%%%%%%%%%%%
%
%   Modifica 8/9/2015 + 18/9/2015
%
%   \item[\emph{(i)}] the pair of cliques $(X, Y)$ with $X = \{x_1, x_2\} \subseteq H$ and $Y = \{y_1, y_2\} \subseteq K$ such that the set $\{x_1, x_2, y_1, y_2\}$ induces a square in $G$ is a semi-homogeneous pair;
%
   \item[\emph{(i)}] let $a_1$ be some node in $A$ which is not universal to $B$ and such that either $a_1 \equiv a_{max}$ or $N(a_{max}) \cap B = B \setminus \{\bar b\}$ and $a_1 \bar b \in E$. Then $(\bar A, \bar B)$ with $\bar A = \{a_1\}$ and $\bar B = B \setminus N(a_1)$ is a semi-homogeneous pair and, for any $b \in \bar B$, $\{a_1, b\}$ is a diagonal of some square in $G$;
   \item[\emph{(ii)}] let $\bar a \in A$ and $\bar B \subset B \setminus N(\bar a)$ be such that $\{\bar a, b\}$ defines a diagonal of some square contained in $A \cup B$ for any $b \in \bar B$. Let $\bar E = \{\bar a b : b \in \bar B\}$; the graph $G'(V, E \cup \bar E)$ is claw-free and $(A, B)$ is a square-semi-homogeneous pair of cliques also in $G'$.
%
%%%%%%%%%%%%%%%%%%%%%%%%%%%%%%%%%%%%%%%%%%%%%%%%%%%
%
\end{description}
\end{Theorem}

\noindent
\emph{Proof}. Let $(\bar A, \bar B)$ be a pair of cliques as in \emph{(i)}. We first prove that $(\bar A, \bar B)$ is a semi-homogeneous pair of cliques. If $\bar B$ contains a single node $b_1$ this is trivially true. Hence assume $|\bar B| \ge 2$ and let $b_1, b_2$ be any pair of nodes in $\bar B$. Assume, by contradiction, that there exists some node $z \in V \setminus (\bar A \cup \bar B)$ which is (without loss of generality) adjacent to $b_2$ and non-adjacent to $a_1$ and $b_1$.

\smallskip\noindent
If $a_1 \equiv a_{max}$ then let $a_2 \in A$ be any node adjacent to $b_2$ (it exists, since $b_2$ is not null to $A$) and let $b_3 \in B$ be a node adjacent to $a_1$ and non-adjacent to $a_2$ (it exists, since $a_1 \in A$ has the maximum number of adjacent nodes in $B$ and $b_2 \in B$ is adjacent to $a_2$ and non-adjacent to $a_1$).

\smallskip\noindent
If, conversely, $a_1 \not\equiv a_{max}$ we have that $N(a_{max}) \cap B = B \setminus \{\bar b\}$ and $a_1 \bar b \in E$. In this case let $a_2 \equiv a_{max}$ and $b_3 \equiv \bar b$. Note that $a_2 b_2 \in E$.

\smallskip\noindent
Observe that, in both cases, $z$ is non-adjacent to $b_3$ (otherwise $(b_3: a_1, b_1, z)$ is a claw in $G$). The set $\{a_1, b_3, b_2, a_2\}$ induces a square in $A \cup B$ and hence, by assumption, the cliques $\{a_1, a_2\}$ and $\{b_2, b_3\}$ define a homogeneous pair. Since $z$ is adjacent to $b_2$ and non-adjacent to $b_3$, it must be null to $\{a_1, a_2\}$. But then $(b_2: b_3, a_2, z)$ is a claw in $G$, a contradiction. It follows that $(\bar A, \bar B)$ is semi-homogeneous.

\medskip\noindent
To complete the proof of property~\emph{(i)} we have to show that any pair $\{a_1, b_1\}$ with $b_1 \in \bar B$ is a diagonal of some square in $G$. If $a_1 \equiv a_{max}$ then let $a_2 \in A$ be any node adjacent to $b_1$ (it exists, since $b_1$ is not null to $A$); if $a_1 \not\equiv a_{max}$ we have that $N(a_{max}) \cap B = B \setminus \{\bar b\}$ and $a_1 \bar b \in E$. In the first case the node $a_2$ has at most as many adjacent nodes in $B$ as the node $a_1$; since $b_1$ is adjacent to $a_2$ and non-adjacent to $a_1$, we have that there exists some node $b_2 \in B$ which is adjacent to $a_1$ and non-adjacent to $a_2$. In the second case, let $a_2 \equiv a_{max}$ and $b_2 \equiv \bar b$. It follows that, in both cases, $\{a_1, a_2, b_1, b_2\}$ induces a square in $G$ as claimed.

\medskip\noindent
To prove property~\emph{(ii)} assume that $\{a_1, a_2, b_1, b_2\}$ induces a square in $G$ with $a_1, a_2 \in A$ and $b_1, b_2 \in B$ and that the edge $a_1 b_1 \in \bar E$ is contained in some claw in $G'$. We claim that this leads to a contradiction and to prove it we only use the property that any edge in $\bar E$ is a diagonal of some square in $G$ (and not the fact that one of the endpoints is necessarily $\bar a$). Hence, without loss of generality, we can assume that the claw in $G'$ is $(a_1: b_1, z_1, z_2)$. Since $z_1$ and $z_2$ are non-adjacent to $b_1$ they belong to $V \setminus B$, so the edges $a_1 z_1$ and $a_1 z_2$ do not belong to $\bar E$. Analogously, the pairs $\{a_2, z_1\}$ and $\{a_2, z_2\}$ do not belong to $\bar E$. If $z_1$ and $z_2$ are both adjacent to $a_2$ in $G$, then $(a_2: z_1, z_2, b_1)$ is a claw in $G$, a contradiction. It follows that either $z_1$ or $z_2$ is not adjacent to $a_2$ in both $G$ and $G'$. Assume (without loss of generality) $a_2 z_2 \notin E$. It follows that $z_2$ does not belong to $A$ and is neither null nor universal to $\{a_1, a_2\}$. Since $z_2 b_1 \notin E$, $z_2$ is not universal to $\{b_1, b_2\}$ and hence, by the assumption that $(A, B)$ is a square-semi-homogeneous pair of cliques in $G$, it must be null to $\{b_1, b_2\}$ in $G$. Consequently $z_2 b_2 \notin E$ and $(a_1: a_2, b_2, z_2)$ is a claw in $G$, a contradiction. Hence $G'$ is claw-free.

\medskip\noindent
Finally, we prove that $(A, B)$ is a square-semi-homogeneous pair of cliques also in $G'$. Assume by contradiction that some diagonal $\bar a b_1 \in \bar E$ of a square $(\bar a, a_2, b_1, b_2)$ in $G$ belongs to a new square $(\bar a, b_1, z_1, z_2)$ in $G'$ with $z_1 \in B$ and $z_2 \in A$ such that the cliques $\{\bar a, z_2\}$ and $\{b_1, z_1\}$ do not define a semi-homogeneous pair. Observe that the edges $\bar a z_2$, $z_2 z_1$ and $z_1 b_1$ do not belong to $\bar E$ (the first because both the endpoints belong to $A$ and the other two because they do not have $\bar a$ as an endpoint). It follows that there exists a node $z \notin A \cup B$ whose neighbors in $\{\bar a, b_1, z_1, z_2\}$ are either $z_1, z_2$ or $\bar a, b_1$. Observe that $\{z_1, z_2\} \cap \{a_2, b_2\} = \emptyset$, since $a_2$ and $b_2$ are adjacent to both $\bar a$ and $b_1$. Suppose first that $z$ is adjacent to $z_1, z_2$ and non-adjacent to $\bar a, b_1$. The node $z$ is not adjacent to $a_2$, otherwise $(a_2: b_1, \bar a, z)$ would be a claw in $G$. Analogously, $b_2 z \notin E$. Moreover, $a_2 z_1 \notin E$ (otherwise $(z_1: a_2, b_2, z)$ would be a claw in $G$). But then $(z_1, z_2, a_2, b_1)$ is a square in $G$ with the property that the cliques $\{a_2, z_2\}$ and $\{b_1, z_1\}$ do not define a semi-homogeneous pair, a contradiction. It follows that $z$ is adjacent to $\bar a, b_1$ and non-adjacent to $z_1, z_2$. The node $z$ is either adjacent to $a_2$ or to $b_2$ (otherwise $(b_1: a_2, b_2, z)$ would be a claw in $G$). Assume, without loss of generality, $z a_2 \in E$. It follows $a_2 z_1 \notin E$, otherwise $(a_2: b_1, z_1, z_2, \bar a, z)$ is a $5$-wheel in $G$, contradicting Lemma~\ref{RegularityLemma}. But then $(b_1, z_1, z_2, a_2)$ is a square in $G$ with the property that the cliques $\{a_2, z_2\}$ and $\{z_1, b_1\}$ do not define a semi-homogeneous pair, a contradiction. Hence $(A, B)$ is a square-semi-homogeneous pair of cliques also in $G'$ and Property~\emph{(ii)} follows.
\done

\begin{Definition}\label{CliqueStrip}
A connected graph $G(V, E)$ is a \emph{clique-strip} if there exists a partition $\{H_0, \dots, H_p\}$ of $V$ in cliques (not necessarily maximal) with the property that $H_i$ is adjacent to $H_j$ only if $|i-j| = 1$ $i, j \in \{0, 1, \dots, p\}$. The clique-strip is said to be \emph{defined} by $\{H_0, \dots, H_p\}$.
\done
\end{Definition}

\begin{Lemma}\label{ShortCliqueStripLemma}
Let $G(V, E)$ be a connected \{claw, net\}-free graph, $Q$ a dominating bisimplicial clique in $G$. Then the set $P = V \setminus N[Q]$ is a clique.
\end{Lemma}

\noindent
\emph{Proof}. Since $Q$ is bisimplicial, its neighborhood is partitioned into two cliques $K_1$ and $K_2$. Let $uv$ be an edge in $E$ with $u \in K_1$ and $v \in K_2$. Assume by contradiction that there exist two non-adjacent nodes $x, y \in P$. Since $Q$ is dominating, the set $P$ is contained in $N(\{u, v\})$. If both $x$ and $y$ were adjacent to $u$ we would have the claw $(u: x, y, z)$ where $z \in Q$ is a node adjacent to $u$. It follows that, without loss of generality, we can assume that $x$ is adjacent to $u$ and non-adjacent to $v$ while, symmetrically, $y$ is adjacent to $v$ and non-adjacent to $u$. Let $z$ be any node in $Q$ adjacent to $u$ and assume that $z$ is non-adjacent to $v$. But then $(u: z, x, v)$ is a claw in $G$, a contradiction. It follows that each node in $Q \cap N(\{u, v\})$ is adjacent to both $u$ and $v$. But now, if $N(\{u, v\}) \supseteq Q$ then $Q \cup \{u, v\}$ is a clique, contradicting the maximality of $Q$. On the other hand, if $N(\{u, v\}) \not\supseteq Q$, let $t$ be a node in $Q \setminus N(\{u, v\})$ and $z$ a node in $N(\{u, v\}) \cap Q$. Then $(z, u, v: t, x, y)$ is a net in $G$, again a contradiction. Hence, we have that $P$ is a clique.
\done

\begin{Lemma}\label{CliqueStripLemma}
Let $G(V, E)$ be a connected \{claw, net\}-free graph with $\alpha(G) \ge 4$ and assume that $G$ contains a bisimplicial clique $Q$ whose neighborhood is partitioned into the cliques $X$ and $Y$ and which is either dominating or strongly bisimplicial. Then $G - X$ ($G - Y$) is the union of at most two disjoint clique-strips defined by clique families which can be constructed in ${\cal O}(|E|)$ time. Moreover, any pair of consecutive cliques $(K_i, K_{i+1})$ in one of the clique-strips is square-semi-homogeneous in $G$.
\end{Lemma}

\noindent
\emph{Proof}. We have two possible cases: either $X$ is null to $Y$ or not. We first prove the lemma in the latter case.

\medskip\noindent
\emph{Case (a). $X$ is not null to $Y$.}

\smallskip\noindent
Since in this case $Q$ is dominating, by Lemma~\ref{ShortCliqueStripLemma}, $P = V \setminus N[Q]$ is a clique. It follows that $\alpha(G) = 4$ and a canonical stable set $S$ in $G$ is composed of four nodes $q_0 \in Q$, $x_0 \in X$, $y_0 \in Y$ and $p_0 \in P$. Since $P$ is null to $Q$ we have that $G - X$ is a clique-strip defined by $\{Q, Y, P\}$, $G - Y$ is a clique-strip defined by $\{Q, X, P\}$ and both can be constructed in ${\cal O}(|E|)$ time. To complete the proof of the lemma in this case, we will show by contradiction that every pair of consecutive cliques in the two clique-strips (namely $(Q, Y)$ and $(Y, P)$ in $G - X$, $(Q, X)$ and $(X, P)$ in $G - Y$) is square-semi-homogeneous in $G$. Since the four cases are symmetric, we can assume, without loss of generality, that there exists a square $H$ in $Q \cup Y$ which is not semi-homogeneous in $G$. Let $H = (q, q', y' y)$ with $q, q' \in Q$, $y, y' \in Y$, $q y' \notin E$, $q' y \notin E$. Since $H$ is not semi-homogeneous in the claw-free graph $G$, there exists a node $x \in V \setminus (Q \cup Y)$ which is, without loss of generality, adjacent to $q, y$ and non-adjacent to $q', y'$. The node $x$ does not belong to $P$ ($P$ is null to $Q$) and hence belongs to $X$.

\medskip\noindent
\emph{Claim (a1). $p_0 y \notin E$.}

\smallskip\noindent
\emph{Proof.} Assume conversely that $p_0$ is adjacent to $y$. Since $p_0 q \notin E$ ($Q$ is null to $P$) we have $p_0 y' \in E$ (otherwise $(y: q, p_0, y')$ is a claw in $G$). Hence $y_0 \ne y, y'$. Since $p_0 q \notin E$ we have $y_0 q \in E$ (otherwise $(y: y_0, p_0, q)$ is a claw in $G$). Analogously, $y_0 q' \in E$. Now observe that $x y_0 \notin E$. In fact otherwise, since $Q$ is dominating, we would also have $p_0 x \in E$ (so $x \ne x_0$) and $(x: x_0, y_0, p_0)$ would be a claw in $G$. But then we have $p_0 x \in E$ (otherwise $(y: y_0, p_0, x)$ is a claw in $G$), so $x \ne x_0$. Moreover, $q$ is not adjacent to $x_0$ (otherwise $(q: q_0, x_0, y_0)$ is a claw in $G$) and hence $(x: x_0, p_0, q)$ is a claw in $G$, a contradiction.

\noindent
\emph{End of Claim (a1).}

\smallskip\noindent
Since $Q$ is dominating, $P \subseteq N(\{x, y\})$. Since $p_0 y \notin E$, we have $p_0 x \in E$, so $x \ne x_0$. Moreover, $x_0 y \notin E$, otherwise we should have $p_0 \in N(\{x_0, y\})$ which is impossible. But then $(x: x_0, y, p_0)$ is a claw in $G$, a contradiction. Hence, if $X$ is not null to $Y$ the lemma follows.

\noindent
\emph{End of Case (a).}

\medskip\noindent
\emph{Case (b). $X$ is null to $Y$.}

\smallskip\noindent
In this case $Q$ is strongly bisimplicial. Let $G_X$ ($G_Y$) be the connected component containing $X$ ($Y$) of the subgraph obtained by removing the nodes in $Q$ from $G$. Let $X_i$ be the set of nodes at distance $i$ from $X$ in $G_X$ (with $X_0 \equiv X$) and let $p$ be the largest integer such that $X_p \ne \emptyset$. Observe that, since $X$ is null to $Y$ and $X_1 \subset N(X_0)$, $X_1$ does not intersect $Y$ and hence is null to $Q$.

\medskip\noindent
\emph{Claim (b1). For each $i \in \{0, \dots, p\}$, $X_i$ is a clique.}

\smallskip\noindent
\emph{Proof.} The claim is true for $i = 0$. Assume that there exists some integer $t \ge 0$ such that the claim is true for $i \le t$ but false for $i = t+1$ and let $x_1, x_2$ be two non-adjacent nodes in $X_{t+1}$. We first observe that $x_1$ and $x_2$ have no common neighbor in $X_t$. Suppose the contrary, let $z$ be such a node and $\overline z$ a node adjacent to $z$ in $X_{t-1}$ if $t \ge 1$ or in $Q$ if $t = 0$. Observe that in both cases $\overline z$ is not adjacent to $x_1$ or $x_2$. In fact, if $t \ge 1$, $\overline z$ belongs to $X_{t-1}$ which is null to $X_{t+1}$; if $t = 0$, $\overline z$ belongs to $Q$ which is null to $X_{t+1} \equiv X_1$. But then $(z: x_1, x_2, \overline z)$ is a claw in $G$, a contradiction. It follows that $x_1$ and $x_2$ (adjacent to $X_t$) have no common neighbor in $X_t$ and, hence, are not universal to $X_t$.

\smallskip\noindent
We claim that there exists a maximal clique $\hat X \supseteq X_t$ such that $x_1$ and $x_2$ belong to $N(\hat X)$ and have no common neighbor in $\hat X$.

\smallskip\noindent
In fact, if $x_1$ and $x_2$ have no common neighbor universal to $X_t$, $\hat X$ can be any maximal clique in $G$ containing $X_t$. If, conversely, $x_1$ and $x_2$ have a common neighbor universal to $X_t$ then, by claw-freeness, $X_t \subseteq N(x_1) \cup N(x_2)$. Let $q$ be a node in $X_{t-1}$ (or in $Q$ if $t = 0$) adjacent to some node $\bar x \in X_t$ and observe that $q$ is non-adjacent to both $x_1$ and $x_2$. Without loss of generality we can assume that $\bar x$ is adjacent to $x_1$. Let $\bar y$ be any node adjacent to $x_2$ in $X_t$ and observe that $\bar x$ is non-adjacent to $x_2$ and $\bar y$ is non-adjacent to $x_1$. If $q \bar y \notin E$ we have that $(\bar x: q, x_1, \bar y)$ is a claw in $G$, a contradiction. It follows that $q$ is adjacent to all the nodes in $N(x_2) \cap X_t$. A symmetric argument shows that $q$ is also universal to $N(x_1) \cap X_t$ and hence it is universal to $X_t$. Moreover, any node $\bar q$ adjacent to both $x_1$ and $x_2$ is non-adjacent to $q$, otherwise $(\bar q: q, x_1, x_2)$ would be a claw in $G$. Let $\hat X$ be a maximal clique containing $X_t \cup \{q\}$. Since $x_1$ and $x_2$ are not adjacent to $q$, they belong to $N(\hat X)$ and have no common neighbor in $\hat X$.

\smallskip\noindent
Hence, as claimed, there exists a maximal clique $\hat X \supseteq X_t$ such that $x_1$ and $x_2$ belong to $N(\hat X)$ and have no common neighbor in $\hat X$. Now, we claim that there exists a node $y$ in $N(\hat X)$ which is non-adjacent to both $x_1$ and $x_2$. In fact, if $t \ge 1$ and $X_{t-1} \not\subseteq \hat X$ such a node clearly exists in $X_{t-1} \setminus \hat X$. If $X_{t-1} \subseteq \hat X$ and $t \ge 2$ such a node clearly exists in $X_{t-2}$. If $t = 0$, since $Q \cup X_0$ is not a clique and $Q$ is null to $X_1$, there exists at least one node $y \in Q$ which is non-adjacent to both $x_1$ and $x_2$ and not universal to $X_0$ and hence belongs to $N(\hat X)$. We are left with the case $t = 1$ and $X_0 \equiv X \subseteq \hat X$. Suppose that every node in $\hat Q = Q \cap N(\hat X)$ is adjacent to either $x_1$ or $x_2$. If both $x_1$ and $x_2$ belong to $N(\hat Q)$ then $x_1$ and $x_2$ both belong to $Y$, contradicting the assumption that $Y$ is a clique. It follows that one of $x_1, x_2$ is null and the other is universal to $\hat Q$. Without loss of generality, assume that $x_1$ is universal to $\hat Q$. Since $Q$ is a maximal clique, we have that there exists a node $u \in Q \setminus \hat Q$ which is not adjacent to $x_1$. Let $v$ be any node in $\hat Q$ and $h \in X_0 \subseteq \hat X$ be a node adjacent to $v$. But then $(v: u, h, x_1)$ is a claw, a contradiction. It follows that also in this case there exists a node $y$ in $N(\hat X)$ which is non-adjacent to both $x_1$ and $x_2$. Hence $x_1, x_2, y$ are three mutually non-adjacent nodes in $N(\hat X)$ with $x_1, x_2$ distant. But then, by \emph{(ii)} of Proposition~\ref{PropertiesNormalClique} $\hat X$ is normal, contradicting the hypothesis that $G$ (and hence $G_X$) is net-free.

\noindent
\emph{End of Claim (b1).}

\smallskip\noindent
Hence, we have that for each $i \in \{0, \dots, p\}$, $X_i$ is a clique, each node in $X_i$ is adjacent to some node in $X_{i-1}$ for $i \ge 1$ and each node in $X_0$ is adjacent to some node in $Q$. It is easy to check that $G_X$ is a clique-strip defined by the family ${\cal X} = \{X_0 \equiv X, X_1, \dots, X_p\}$. Analogously, also $G_Y$ is a clique-strip, possibly coincident with $G_X$ and defined by the family ${\cal Y} = \{Y_0 \equiv Y, Y_1, \dots, Y_t\}$. Moreover, ${\cal X}$ and ${\cal Y}$ can be constructed in ${\cal O}(|E|)$ time.

\medskip\noindent
If $G_X \not\equiv G_Y$ then $N(Q) \cap Y_i \ne \emptyset$ if and only if $Y_i = Y_0 = Y$. Hence $\{Q\} \cup {\cal Y}$ and ${\cal X} \setminus \{X\}$ are clique families defining two clique-strips which are disjoint, null to each other and partition $G - X$. Let $(K_i, K_{i+1})$ be a pair of consecutive cliques in one of the clique strips; we claim that $(K_i, K_{i+1})$ is square-semi-homogeneous in $G$. In fact, let $C = \{a_1, a_2, b_1, b_2\}$ be any set inducing a square in $K_i \cup K_{i+1}$, with $\{a_1, a_2\} \subseteq K_i$ and $\{b_1, b_2\} \subseteq K_{i+1}$. Let $u$ be any node in $V \setminus C$ which is not null to $C$. We have that $u$ belongs to one of the three sets $K_i \cup K_{i-1}$ ($K_{i-1}$ possibly empty), $K_{i+1} \cup K_{i+2}$ ($K_{i+2}$ possibly empty), $X$. If $u \in K_i \cup K_{i-1}$ we have that $u$ is null to $\{b_1, b_2\}$ and, by claw-freeness, universal to $\{a_1, a_2\}$. If $u \in K_{i+1} \cup K_{i+2}$ we have that $u$ is null to $\{a_1, a_2\}$ and universal to $\{b_1, b_2\}$. Finally, since each node in $X$ can be adjacent only to nodes in $X_1$ or in $Q$, if $u \in X$ we have either $K_i \equiv X_1$ or $K_i \equiv Q$. In both cases $u$ is null to $\{b_1, b_2\}$ and, by claw-freeness, universal to $\{a_1, a_2\}$. It follows that $(\{a_1, a_2\}, \{b_1, b_2\})$ is semi-homogeneous in $G$ and hence $(K_i, K_{i+1})$ is square-semi-homogeneous in $G$. Analogously, ${\cal Y} \setminus \{Y\}$ and $\{Q\} \cup {\cal X}$ are clique families defining two clique-strips which are disjoint, null to each other and partition $G - X$. Moreover, any pair of consecutive cliques $(K_i, K_{i+1})$ in one of the clique strips is square-semi-homogeneous in $G$. Consequently, if $G_X \not\equiv G_Y$ the lemma follows.

\smallskip\noindent
Assume now $G_X \equiv G_Y$. Observe that $X \equiv X_0$ is covered by the cliques in ${\cal Y}$ and intersects at most two consecutive cliques in ${\cal Y}$. Moreover $X_0$ is null to $Y_0$ and has an empty intersection with $Y_1$.

\medskip\noindent
\emph{Claim (b2). $X_0$ can only intersect the cliques $Y_t$ and $Y_{t-1}$. Moreover, if it intersects $Y_{t-1}$ then $Y_t \setminus X_0 = \emptyset$.}

\smallskip\noindent
\emph{Proof.} Let $k \ge 2$ be the smallest index such that $X_0 \cap Y_k \ne \emptyset$ and suppose that either $k \le t-2$ or $k = t-1$ and $Y_t \setminus X_0 \ne \emptyset$. Observe that $Y_k \subseteq (X_0 \cup X_1)$. Moreover, each node in $Y_{k+1}$ is adjacent to some node in $Y_k$ and hence we have $Y_{k+1} \subseteq (X_0 \cup X_1 \cup X_2)$. Since each node in $X_0 \cap Y_k$ is adjacent to some node in $Y_{k-1}$ and $X_0 \cap Y_{k-1} = \emptyset$ by assumption, we have that $X_1 \cap Y_{k-1} \ne \emptyset$ and hence $X_1 \cap Y_{k+1} = \emptyset$. Moreover, $Y_{k-1} \cup Y_{k-2}$ contains some node in $X_2$. It follows that $X_2 \cap Y_{k+1} = \emptyset$. Consequently, $Y_{k+1} \subseteq X_0$. It follows, by assumption, that $k \le t-2$ and $Y_{k+2}$ is non-empty. Since each node in $Y_{k+2}$ is adjacent to some node in $Y_{k+1} \subseteq X_0$ and $X_0 \cap Y_k \ne \emptyset$, we have $Y_{k+2} \subseteq X_1$. Since $X_1 \cap Y_{k-1} \ne \emptyset$ we have the contradiction that $X_1$ is not a clique.

\noindent
\emph{End of Claim (b2).}

\smallskip\noindent
Let ${\cal Y}'$ be the clique family obtained from ${\cal Y}$ by substituting $Y_{t-1}$ and $Y_t$ with $Y_{t-1}' \equiv Y_{t-1} \setminus X$ and $Y_t' \equiv Y_t \setminus X$, respectively, possibly removing $Y_t'$ if empty. We have that $G - X$ is a clique-strip defined by the family $\{Q\} \cup {\cal Y}'$. We claim that any pair of consecutive cliques $(K_i, K_{i+1})$ in $G - X$ is square-semi-homogeneous in $G$. In fact, as above, let $C = \{a_1, a_2, b_1, b_2\}$ with $\{a_1, a_2\} \subseteq K_i$ and $\{b_1, b_2\} \subseteq K_{i+1}$ induce a square and let $u$ be any node in $V \setminus C$ which is not null to $C$. If $u \notin X$, the same argument used above shows that $u$ is universal to $\{a_1, a_2\}$ and null to $\{b_1, b_2\}$ or viceversa. If, on the other hand, $u$ belongs to $X$ then, by Claim~\emph{(b2)}, $u$ belongs to $Y_t \cup Y_{t-1}$ and can be adjacent only to nodes in $Y_{t-2}$, $Y_{t-1}'$, $Y_t'$ or $Q$. Moreover, if $u$ is adjacent to some node in $Y_{t-2}$ then $u$ belongs to $Y_{t-1}$ and $Y_t'$ is empty. It follows that $u$ is either null or universal to at least one of $K_i, K_{i+1}$. Moreover, if $u$ is null to $\{a_1, a_2\}$ ($\{b_1, b_2\}$) then, by claw-freeness, it is universal to $\{b_1, b_2\}$ ($\{a_1, a_2\}$). Hence $(\{a_1, a_2\}, \{b_1, b_2\})$ is semi-homogeneous and $(K_i, K_{i+1})$ is square-semi-homogeneous in $G$. Analogously, $G - Y$ is a clique-strip defined by the family $\{Q\} \cup {\cal X}'$, where ${\cal X}'$ is defined in analogy with ${\cal Y}'$. Moreover, any pair of consecutive cliques $(K_i, K_{i+1})$ in $G - Y$ is square-semi-homogeneous in $G$. Hence, also if $G_X \equiv G_Y$ the lemma follows and we are done.
\done

\begin{Theorem}\label{CliqueStripTheorem}
Let $G(V, E)$ be a connected \{claw, net\}-free graph and let $S = \{s_1, s_2, \dots, s_t\}$ be a canonical stable set of $G$ ($t \ge 4$). Then $G$ contains a clique $X$ such that $G - X$ is the union of at most two clique-strips. Moreover, any pair of consecutive cliques $K_i, K_{i+1}$ in one of the clique-strips is square-semi-homogeneous. The clique $X$ and the clique families defining the clique-strips can be found in ${\cal O}(|E|)$ time.
\end{Theorem}

\noindent
\emph{Proof}. Since $\alpha(G) \ge 4$, by Lemma~\ref{RegularityLemma} $G$ is quasi-line and hence each node $s_i \in S$ is regular. It follows that $N[s_i]$ is covered by two cliques, say $C_{s_i}$ and $\overline C_{s_i}$.
%
%%%%%%%%%%%%%%%%%%%%%%%%%%%%%%%%%%%%%%%%%%%%%%%%%%%%%%%%%%%%%%%%%%%%%
%%%
%%%   Risposta alla seconda obiezione del primo referee.
%%%
%%%%%%%%%%%%%%%%%%%%%%%%%%%%%%%%%%%%%%%%%%%%%%%%%%%%%%%%%%%%%%%%%%%%%
%
\medskip\noindent
Let $H(S, T)$ be the graph where $x y$ is an edge in $T$ if and only if $W(x, y)$ is a non-empty wing in $G$. Observe that, by Theorem~\ref{3WingedNormal}, the degree of each node $u$ in $H$ is at most $2$.

\medskip\noindent
\emph{Claim (i). The graph $H$ is connected.}

\smallskip\noindent
\emph{Proof.} Assume conversely that there exist at least two connected components $C_1, C_2$ of $H$. Let $P = (s_1, z_1, \dots, z_h, s_2)$ be a shortest path in $G$ connecting two nodes $s_1, s_2$ in different components of $H$ (it exists since $G$ is connected). Without loss of generality, assume $s_1 \in C_1$ and $s_2 \in C_2$. By minimality of $P$, $z_i \notin S$ ($i = 1, \dots, h$) and $h \ge 2$, otherwise $W(s_1, s_2)$ would be a wing containing $z_1$, contradicting the assumption that $s_1$ and $s_2$ do not belong to the same connected component of $H$. If $h \ge 3$ we have that there exists at least one node $s_3 \in S$ adjacent to $z_2$ and different from $s_1$ and $s_2$. If $s_3 \notin C_1$ then the path $(s_1, z_1, z_2, s_3)$ contradicts the minimality of $P$. It follows $s_3 \in C_1$ and hence $s_3, s_2$ are in different connected components of $H$. But then $(s_3, z_2, z_3, \dots, z_h, s_2)$ is a path in $G$ connecting two nodes in different components of $H$ which is shorter than $P$, a contradiction. Consequently, we have $h = 2$. Assume that $z_1$ is a bound node. Then there exists a node $s_3 \in S$ adjacent to $z_1$ and different from $s_1$ and $s_2$. By claw-freeness $s_3$ is also adjacent to $z_2$. It follows that both $W(s_1, s_3)$ and $W(s_2, s_3)$ are non-empty wings and hence $s_3$ belongs to both $C_1$ and $C_2$, a contradiction. The same argument applies if $z_2$ is a bound node and hence both $z_1$ and $z_2$ are free nodes. But then $W(s_1, s_2)$ is non-empty, again a contradiction.

\noindent
\emph{End of Claim (i).}

\smallskip\noindent
Since $|S| \ge 4$, by the previous claim $H$ is either a path with at least three edges or a cycle with at least four edges so we can assume without loss of generality that each node $s_i \in S$ ($2 \le i \le t-1$) defines non-empty wings only with $s_{i-1}$ and $s_{i+1}$. Moreover, if $H$ is a cycle then also the wing $W(s_t, s_1)$ is non-empty. We now prove a claim on the adjacency structure of $G$.

\medskip\noindent
\emph{Claim (ii). A node in $N[s_i]$ is only adjacent to nodes in $N[s_{i-1}] \cup N[s_i] \cup N[s_{i+1}]$ ($i \in \{2, \dots, t-1\}$).}

\smallskip\noindent
\emph{Proof.} Let $u$ be any node in $N[s_i]$. Since $s_i$ defines wings only with $s_{i-1}$ and $s_{i+1}$, we have that $N(u) \cap S \subseteq \{s_{i-1}, s_i, s_{i+1}\}$. Assume now, by contradiction, that there exists a node $z \notin N[s_{i-1}] \cup N[s_i] \cup N[s_{i+1}]$ adjacent to $u$ (note that this implies that $u$ is not $s_i$ and $z$ is not a stable node). If $z$ is a bound node adjacent to $s_j, s_k \in S \setminus \{s_{i-1}, s_i, s_{i+1}\}$ we have that $(z: s_j, s_k, u)$ is a claw in $G$, a contradiction. If $u$ is bound suppose, without loss of generality, that it is adjacent to $s_{i-1}$. But then $(u: s_{i-1}, s_i, z)$ is a claw in $G$, again a contradiction. It follows that both $u$ and $z$ are free. But then the wing $W(s_i, S(z))$ is non-empty, contradicting the assumption that $s_i$ defines wings only with $s_{i-1}$ and $s_{i+1}$.

\noindent
\emph{End of Claim (ii).}
%
%%%%%%%%%%%%%%%%%%%%%%%%%%%%%%%%%%%%%%%%%%%%%%%%%%%%%%%%%%%%%%%%%%%%%
%%%
%%%   Risposta alla prima obiezione del secondo referee: "... Hence
%%%   in the same sentence, the maximal clique contains W(s1,s2) cap
%%%   N(s2), not W(s2,s3) cap N(s2), so Claim (ii) cannot be applied."
%%%   Il Claim (ii) viene modificato dimostrando che il risultato vero
%%%   per W(s2, s3) vale anche per W(s1, s2).
%%%
%%%%%%%%%%%%%%%%%%%%%%%%%%%%%%%%%%%%%%%%%%%%%%%%%%%%%%%%%%%%%%%%%%%%%
%

\medskip\noindent
\emph{Claim (iii). Let $i \in \{2, \dots, t-1\}$ and let $Q$ be a maximal clique in $N[s_i]$. Suppose $Q$ contains $s_i$ and either \emph{(a)} $W(s_{i-1}, s_i) \cap N(s_i)$ or \emph{(b)} $W(s_i, s_{i+1}) \cap N(s_i)$. Then $Q$ is a bisimplicial clique which is either dominating or strongly bisimplicial.}

\smallskip\noindent
\emph{Proof.} By Claim~\emph{(ii)}, $N[Q]$ is contained in $N[s_{i-1}] \cup N[s_i] \cup N[s_{i+1}]$. We partition the neighborhood of $Q$ into two sets $X$ and $Y$. In particular, in case~\emph{(a)} we let  $X = N(Q) \cap (N[s_i] \cup N[s_{i+1}])$ and, in case~\emph{(b)}, we let  $X = N(Q) \cap (N[s_{i-1}] \cup N[s_i])$. Observe that case~\emph{(a)} with $i = j$ is symmetric to case~\emph{(b)} with $i = t-j+1$ for any $j \in \{2, \dots, t-1\}$. Hence, without loss of generality, in what follows we consider only case~\emph{(a)} and let $X = N(Q) \cap (N[s_i] \cup N[s_{i+1}])$ and $Y = N(Q) \cap N[s_{i-1}]$. Since $N[s_{i-1}] \cap N[s_{i+1}]$ is empty and $N[s_{i-1}] \cap N[s_i]$ is contained in $Q$ (by assumption \emph{(a)}) we have that $(X, Y)$ is a partition of $N(Q)$; moreover any node in $X$ belongs to $N(s_i)$ or to $N[s_{i+1}]$ but not to $N[s_{i-1}]$ and any node in $Y$ belongs to $N[s_{i-1}]$ but not to $N[s_i] \cup N[s_{i+1}]$.

\smallskip\noindent
Observe that both $X$ and $Y$ are non-empty. In fact, since $W(s_{i-1}, s_i)$ is non-empty, we have that $N(Q)$ contains either $s_{i-1}$ or a free node in $N(s_{i-1})$ and hence we have $Y \ne \emptyset$. On the other hand, if $N(s_i) \setminus Q$ is non-empty then $X$ is non-empty, while if $N(s_i) \subset Q$ then the assumption that $W(s_i, s_{i+1})$ is non-empty implies that either $s_{i+1}$ or a free node in $N(s_{i+1})$ is adjacent to $Q$ and belongs to $X$.

\smallskip\noindent
Assume first that $X$ is null to $Y$. If $X$ is not a clique, let $x_1, x_2$ be two non-adjacent nodes in $X$ and let $y$ be any node in $Y$. The nodes $x_1$, $x_2$ and $y$ are three mutually non-adjacent nodes in $N(Q)$, contradicting Theorem~\ref{NotReducibleIsNormal}. It follows that $X$ is a clique and, by a similar argument, $Y$ is a clique. As a consequence, $Q$ is strongly bisimplicial and Claim~\emph{(iii)} follows.

\smallskip\noindent
Assume now that $X$ is not null to $Y$ and let $x \in X$ and $y \in Y$ be any pair of adjacent nodes. We have the following:

\medskip\noindent
\emph{Claim (iii.1). The stable set $S$ contains exactly four nodes and there exists a square $C = (x, y, y', x')$ with $N(y') \cap S = \{s_{i-1}, s_i\}$, $N(x') \cap S = \{s_i, s_{i+1}\}$, $N(x) \cap S = \{s_{i+1}, s_{i+2}\}$ and $N(y) \cap S = \{s_{i-1}, s_{i+2}\}$ (sums taken modulo $4$).}

\smallskip\noindent
\emph{Proof.} Observe first that $x$ is not stable (otherwise we would have $x \equiv s_{i+1}$, contradicting $y \notin N[s_{i+1}]$) and $y$ is not stable (otherwise we would have $y \equiv s_{i-1}$, contradicting $x \notin N[s_{i-1}]$). Moreover, $x$ and $y$ are not both free (otherwise $x$ would either belong to $W(s_{i-1}, s_{i+1})$, which is empty, or to $(W(s_{i-1}, s_i) \cap N(s_i)) \setminus Q$, which is also empty).

\smallskip\noindent
If $x$ is free and $y$ is bound, then we have $x \in F(s_i) \cup F(s_{i+1})$ and $y \in W(s_{i-1}, s_{i-2})$ (in this case the wing $W(s_{i-1}, s_{i-2})$ is non-empty). But then $(y: x, s_{i-1}, s_{i-2})$ is a claw in $G$, a contradiction.

\smallskip\noindent
If $x$ is bound and $y$ is free, then we have $y \in F(s_{i-1})$ and either $x \in W(s_i, s_{i+1})$ or $x \in W(s_{i+1}, s_{i+2})$. In the first case $(x: y, s_i, s_{i+1})$ is a claw in $G$; in the second case $(x: y, s_{i+1}, s_{i+2})$ is a claw in $G$. In both cases we have a contradiction.

\smallskip\noindent
It follows that both $x$ and $y$ are bound and $y$ belongs to $W(s_{i-1}, s_{i-2})$. On the other hand, $x \notin W(s_i, s_{i+1})$ (otherwise $(x: y, s_i, s_{i+1})$ would be a claw in $G$) and hence $x \in W(s_{i+1}, s_{i+2})$. By claw-freeness, we have $s_{i-2} \equiv s_{i+2}$, otherwise $(y: x, s_{i-1}, s_{i-2})$ would be a claw in $G$. It follows that $t = 4$ and that $W(s_1, s_4)$ is non-empty.

\smallskip\noindent
Now assume that $x$ and $y$ have a common neighbor $z \in Q$. Since $x$ and $y$ are not adjacent to $s_i$, we have $z \ne s_i$ and hence $(x, y, z: s_{i+1}, s_{i-1}, s_i)$ is a net in $G$, a contradiction. It follows that there exist nodes $x', y' \in Q$ such that $x$ is adjacent to $x'$ and non-adjacent to $y'$ while $y$ is adjacent to $y'$ and non-adjacent to $x'$. Moreover, $x'$ is adjacent to $s_{i+1}$ (otherwise $(x: s_{i+1}, x', s_{i+2})$ would be a claw in $G$) and $y'$ is adjacent to $s_{i-1}$ (otherwise $(y: s_{i-1}, y', s_{i+2})$ would be a claw in $G$). It follows that $C = \{x, y, y', x'\}$ induces a square in $G$, with $x \in W(s_{i+1}, s_{i+2})$, $y \in W(s_{i-1}, s_{i+2})$, $y' \in W(s_{i-1}, s_i)$ and $x' \in W(s_i, s_{i+1})$. The claim follows.

\noindent
\emph{End of Claim (iii.1).}

\medskip\noindent
\emph{Claim (iii.2). The square $C$ dominates $V$ ($N[C] = V$).}

\smallskip\noindent
\emph{Proof.} Suppose conversely that there exists a node $z \notin N[C]$. Observe that $C_1 = C \cup \{s_{i-1}, s_i\}$ induces a subgraph isomorphic to $S_3^-$ and hence, by Lemma~\ref{Brandstadt}, every node in $V \setminus C_1$ is adjacent to two nodes in $C_1$. Since $z$ is null to $C$, we have that it is adjacent to both $s_{i-1}$ and $s_i$. Analogously, $C_2 = C \cup \{s_i, s_{i+1}\}$ induces a subgraph isomorphic to $S_3^-$ and hence $z$ is adjacent to both $s_i$ and $s_{i+1}$. But then $(z: s_{i-1}, s_i, s_{i+1})$ is a claw in $G$, a contradiction. It follows that $N[C] = V$ as claimed.

\noindent
\emph{End of Claim (iii.2).}

\medskip\noindent
\emph{Claim (iii.3). Each node $z \in V$ which is non-adjacent to both $x$ and $y$ belongs to $Q$.}

\smallskip\noindent
\emph{Proof.} Observe that we have $z \notin C$ and, since $N[C] = V$ and $G$ is claw-free, $z$ is adjacent to both $x'$ and $y'$. Suppose, by contradiction, that $z$ does not belong to $Q$ and hence is not adjacent to some node $\overline z \in Q$. It follows that $\overline z$ is adjacent to $x$ (otherwise $(x': z, \overline z, x)$ would be a claw in $G$) and to $y$ (otherwise $(y': z, \overline z, y)$ would be a claw in $G$). Consequently, $\overline z$ is not $s_i$, is universal to $C$ and adjacent to $s_i$ (since $\overline z$ belongs to $Q$). The assumption that $W(s_i, s_{i+2})$ is empty implies that $\overline z$ is not adjacent to $s_{i+2}$ and hence is adjacent to $s_{i-1}$ (otherwise $(y: s_{i-1}, s_{i+2}, \overline z)$ would be a claw). But then $(\overline z: s_{i-1}, s_i, x)$ is a claw in $G$, a contradiction.

\noindent
\emph{End of Claim (iii.3).}

\medskip\noindent
\emph{Claim (iii.4). $X$ and $Y$ are cliques.}

\smallskip\noindent
\emph{Proof.} Assume, by contradiction, that $X$ is not a clique and let $x_1, x_2$ be two non-adjacent nodes in $X$. If $x_1$ and $x_2$ are not adjacent to $y$, then $x_1$, $x_2$ and $y$ are three mutually non-adjacent nodes in $N(Q)$, contradicting Theorem~\ref{NotReducibleIsNormal}. It follows that either $x_1$ or $x_2$ is adjacent to $y$. Suppose $x_1 y \in E$ and, without loss of generality, let $x_1 \equiv x$. By Claim~\emph{(iii.3)} we have that also $x_2$ must be adjacent to $y \in N[s_{i-1}]$. Since $x_1$ and $x_2$ are not adjacent to $s_{i-1}$, we have $y \ne s_{i-1}$ and $(y: x_1, x_2, s_{i-1})$ is a claw in $G$, a contradiction. It follows that $X$ is a clique. A symmetric argument shows that also $Y$ is a clique.

\noindent
\emph{End of Claim (iii.4).}

Hence we have proved that when $X$ is not null to $Y$ $Q$ is a bisimplicial dominating clique and case~\emph{(a)} of Claim~\emph{(iii)} follows. The proof of case~\emph{(b)} is symmetric.

\noindent
\emph{End of Claim (iii).}

%
%%%%%%%%%%%%%%%%%%%%%%%%%%%%%%%%%%%%%%%%%%%%%%%%%%%%%%%%%%%%%%%%%%%%%
%%%
%%%   Risposta alla prima obiezione del secondo referee: "... Hence
%%%   in the same sentence, the maximal clique contains W(s1,s2) cap
%%%   N(s2), not W(s2,s3) cap N(s2), so Claim (ii) cannot be applied."
%%%   Riga +5: (namely $W(s_1, s_2)$) --> (namely $W(s_2, s_3)$)
%%%   Riga +6: $W(s_2, s_3) \cap N(s_2)$ --> $W(s_1, s_2) \cap N(s_2)$
%%%   Riga +6: $\overline C_{s_2}$ is a strongly bisimplicial clique
%%%     --> $\overline C_{s_2}$ is a dominating or strongly bisimplicial clique
%%%
%%%%%%%%%%%%%%%%%%%%%%%%%%%%%%%%%%%%%%%%%%%%%%%%%%%%%%%%%%%%%%%%%%%%%
%

\medskip\noindent
Let $A = C_{s_2} \cap W(s_1, s_2)$, $\overline A = \overline C_{s_2} \cap W(s_1, s_2)$, $B = C_{s_3} \cap W(s_3, s_4)$ and $\overline B = \overline C_{s_3} \cap W(s_3, s_4)$. The sets $C_{s_2}, \overline C_{s_2}, C_{s_3}, \overline C_{s_3}, A, \overline A, B, \overline B$ can be constructed in ${\cal O}(|N(s_2) \cup N(s_3)|^2) = {\cal O}(|E|)$. If $A$ is empty then $W(s_1, s_2) \cap N(s_2)$ is contained in $\overline C_{s_2}$. Then, by Claim~\emph{(iii)}, $\overline C_{s_2}$ is a dominating or strongly bisimplicial clique. In this case, by Lemma~\ref{CliqueStripLemma}, we are done. Hence, we can assume that $A$ is non-empty and, analogously, also $\overline A, B, \overline B$ are non-empty.

\medskip\noindent
\emph{Claim (iv). If $A, \overline A, B, \overline B$ are all non-empty then $W(s_2, s_3) \cap N(s_2)$ is a clique.}

\smallskip\noindent
\emph{Proof.} Suppose that $W(s_2, s_3) \cap N(s_2)$ is not a clique. Let $x \in C_{s_2}$ and $y \in \overline C_{s_2}$ be two non-adjacent nodes in $W(s_2, s_3) \cap N(s_2)$. Observe that, since $S$ is a canonical stable set, either $x$ and $y$ are both bound nodes or one is bound and the other is free. Assume, without loss of generality, that $x$ is bound and belongs to $C_{s_3}$. Let $t_1$ be a node in $A$ and $t_2$ a node in $B$. Observe that both $t_1$ and $t_2$ are adjacent to $x$ and that $t_1$ belongs to $W(s_1, s_2)$. If $t_1$ is bound let $\overline t_1 \equiv s_1$ while if $t_1$ is free let $\overline t_1$ be any free node in $F(s_1) \cap N(t_1)$. In both cases, $\overline t_1$ is null to $N[s_3]$ ($W(s_1, s_3) = \emptyset$). Assume $\overline t_1 y \in E$. We have that either $(y: \overline t_1, s_2, s_3)$ (if $y$ is bound) or $(y: \overline t_1, s_2, u)$ (if $y$ is free and $u$ is any node in $F(s_3) \cap N(y)$) is a claw in $G$, a contradiction. It follows $\overline t_1 y \notin E$. Moreover, $\overline t_1$ is not adjacent to $t_2$, for otherwise $W(s_1, s_3)$ would be non-empty. Analogously, if $t_2$ is bound let $\overline t_2 \equiv s_4$ and if $t_2$ is free let $\overline t_2$ be any free node in $F(s_4) \cap N(t_2)$. In both cases, if $t_2$ is adjacent to $y$ we have that $(t_2: \overline t_2, x, y)$ is a claw in $G$, a contradiction. It follows that $t_2$ is not adjacent to $y$. But then $\overline t_1$, $y$ and $t_2$ are three mutually non-adjacent nodes in $N(C_{s_2})$ implying that $C_{s_2}$ is not reducible, contradicting Theorem~\ref{NotReducibleIsNormal}.

\noindent
\emph{End of Claim (iv).}

\medskip\noindent
Now, observe that the set $W = (W(s_2, s_3) \cap N(s_2)) \cup \{s_2\}$ can be constructed in ${\cal O}(|N(s_2) \cup N(s_3)|^2) = {\cal O}(|E|)$ time. Moreover, a maximal clique $Q$ containing $W$ can also be constructed in ${\cal O}(|E|)$ time (recall that the size of $N(v)$ is ${\cal O}(\sqrt{|E|})$ for any node $v \in V$). Hence, the theorem follows by Claim~\emph{(iii)} and by Lemma~\ref{CliqueStripLemma}.
\done

\section{The Maximum Weight Stable Set Problem on \{claw, net\}-free graphs}
%===========================================================================

\noindent
In this section we reduce the maximum weight stable set problem in a node-weighted \{claw, net\}-free graph $G(V, E)$ to $\sqrt{|E|} + 1$ maximum weight stable set problems in suitably defined \emph{interval graphs}. In \cite{Olariu91} Olariu proved that a graph is an interval graph if and only if its nodes admit a \emph{consistent ordering}, defined as follows.

\begin{Definition}\label{Consistency}
Let $G(V, E)$ be a connected graph. An ordering $\{v_1, v_2, \dots, v_n\}$ of $V$ is said to be \emph{consistent} if each triple $\{i, j, k\}$ with $1 \le i < j < k \le n$ and $v_i v_k \in E$ satisfies $v_j v_k \in E$.
\done
\end{Definition}

\medskip\noindent
We first prove that a claw-free square-free clique-strip admits a consistent ordering (which can be found in ${\cal O}(|E| + |V|\log|V|)$ time) and hence is an interval graph.

\begin{Lemma}\label{ConsistentOrdering}
Let $G(V, E)$ be a connected claw-free square-free clique-strip defined by the family of cliques $\{K_1, K_2, \dots, K_p\}$. Then for any pair of nodes $v_h, v_k \in K_t$ ($t \in \{1, \dots, p-1\}$) either $N(v_h) \cap K_{t+1} \subseteq N(v_k) \cap K_{t+1}$ or $N(v_k) \cap K_{t+1} \subseteq N(v_h) \cap K_{t+1}$. Moreover an ordering $\{v_1, v_2, \dots, v_n\}$ of $V$ with $v_h \prec v_k$ if:

\begin{description}
   \item[\emph{(i)}] $v_h \in K_t$ and $v_k \in K_l$ for some $1 \le t < l \le p$; or
   \item[\emph{(ii)}] $v_h, v_k \in K_t$ for some $t \in \{1, \dots, p-1\}$ and $N(v_h) \cap K_{t+1} \subseteq N(v_k) \cap K_{t+1}$;
\end{description}

\noindent
always exists, is consistent and can be found in ${\cal O}(|E| + |V|\log|V|)$ time.
\end{Lemma}

\noindent
\emph{Proof}. Assume that there exists a pair of nodes $v_h, v_k$ in some clique $K_t$ ($t < p$) such that neither $N(v_h) \cap K_{t+1} \subseteq N(v_k) \cap K_{t+1}$ nor $N(v_k) \cap K_{t+1} \subseteq N(v_h) \cap K_{t+1}$. Let $\bar v_h$ be a node in $K_{t+1}$ which is non-adjacent to $v_h$ and adjacent to $v_k$. Analogously, let $\bar v_k$ be a node in $K_{t+1}$ which is non-adjacent to $v_k$ and adjacent to $v_h$. Then, the four nodes $\{v_h, \bar v_k, \bar v_h, v_k\}$ induce a square in $G[K_t \cup K_{t+1}]$, a contradiction. Now, let $\{v_1, v_2, \dots, v_n\}$ be an ordering of $V$ with $v_h \prec v_k$ if condition \emph{(i)} or \emph{(ii)} is satisfied. Suppose that the ordering is not consistent and let $v_i, v_j, v_k$ be three nodes such that $1 \le i < j < k \le |V|$, $v_i v_k \in E$ and $v_j v_k \notin E$. If $v_i, v_k$ belong to the same clique $K_t$ then, by property \emph{(i)}, also $v_j$ belongs to $K_t$, contradicting the assumption that $v_j v_k \notin E$. Hence, without loss of generality, we can assume $v_i \in K_t$, $v_k \in K_{t+1}$ and $v_j \in K_t \cup K_{t+1}$ for some $t \in \{1, \dots, p-1\}$. Since $v_j v_k \notin E$, we have $v_j \in K_t$. Moreover, since $v_j \not\prec v_i$, by property \emph{(ii)} we have $N(v_j) \cap K_{t+1} \not\subseteq N(v_i) \cap K_{t+1}$ and hence $N(v_i) \cap K_{t+1} \subset N(v_j) \cap K_{t+1}$. But this contradicts the assumption that $v_i v_k \in E$ and $v_j v_k \notin E$ and the lemma follows.
\done

% We recall that a pair of cliques $(H, K)$ is said to be \emph{proper} if each node in $H$ ($K$) is neither null nor universal to $K$ ($H$).

\medskip\noindent
%
%%%%%%%%%%%%%%%%%%%%%%%%%%%%%%%%%%%%%%%%%%%%%%%%%%%
%
%   Modifica 7/9/2015
%
In the previous section we have shown that a connected \{claw, net\}-free graph $G$ with $\alpha(G) \ge 4$ contains a clique $X$ such that $G - X$ is decomposed in (at most) two clique-strips with the property that each pair of consecutive cliques in the strips is square-semi-homogeneous in $G$. We now show how to remove all the squares in the clique-strips decomposing $G - X$ while preserving at least one maximum weight stable set of the original graph (with respect to some node-weighting vector $w$). This task will be accomplished by Procedure~\emph{Interval($G$)} whose pseudo-code can be found below.
\medskip\noindent
Let $\{K_1, K_2, \dots, K_p\}$ be a clique family defining one of the strips. Since any square is contained in a pair of consecutive (square-semi-homogeneous) cliques of the family, the core operation of this procedure is the addition of edges having the end-points in such pairs. In particular, for each pair $K_i$ and $K_{i+1}$ (\emph{inner loop}), we define the two cliques $A = K_i \cap N(K_{i+1})$ and $B = K_{i+1} \cap N(K_i)$ and for each square induced in $A \cup B$, we add at least one edge having as end-points the nodes of a diagonal. Observe that any square in $K_i \cup K_{i+1}$ is contained in $A \cup B$ and that the cliques $A$ and $B$ satisfy the hypotheses of Theorem~\ref{C4Properties}. Hence, by Property~\emph{(ii)}, such addition does not produce new claws and preserves the property that $(A, B)$ is a square-semi-homogeneous pair of cliques.

\medskip\noindent
The inner loop iterates through stages. At each stage, either \emph{(i)} a node $a \in A$ universal to $B$ is removed from $A$ along with any node in $B$ null to $A \setminus \{a\}$; or \emph{(ii)} an edge $a_1 b_1$ is added which is a diagonal of some square $(a_1, a_2, b_1, b_2)$ with $a_1, a_2 \in A$, $b_1, b_2 \in B$ and with the property that $N(a_1) \cap B = B \setminus \{b_1\}$ and that $w(a_1) + w(b_1) \ge w(a_2) + w(b_2)$; or \emph{(iii)} for some node $\bar a \in A$ all the edges $\bar a b$ are added with $b \in B \setminus (N(\bar a) \cup \{b^*\})$, where $b^*$ is the node in $B \setminus N(\bar a)$ having maximum weight.

\medskip\noindent
The inner loop maintains throughout an integer vector $d[\cdot]$ whose entries are associated with the nodes in $A \cup B$ and such that for each $a \in A$ ($b \in B$) $d[a] = |N(a) \cap B|$ ($d[b] = |N(b) \cap A|$). The initialization of the vector $d[\cdot]$ can be done in ${\cal O}(|K_i| |K_{i+1}|)$ time. At each stage let $a_{max} \in A$ be the node maximizing $d[\cdot]$ in $A$; such a node can be found in time ${\cal O}(|K_i|)$.

\medskip\noindent
If $d[a_{max}] = |B|$ then case~\emph{(i)} applies and $a_{max}$ is removed from $A$, $d[\cdot]$ is updated for the nodes in $B$ and any node $b \in B$ with $d[b] = 0$ is also removed (procedure \emph{Remove}$(a_{max})$). This task can be accomplished in time ${\cal O}(|K_{i+1}|)$.

\medskip\noindent
If $d[a_{max}] = |B| - 1$ we let $b_1$ be the unique node in $B$ which is not adjacent to $a_{max}$ and $a_2$ be any node in $A$ adjacent to $b_1$. If also $d[a_2] = |B| - 1$ then Case~\emph{(ii)} applies. In fact, letting $b_2$ be the unique node in $B$ which is not adjacent to $a_2$, we have the square $(a_{max}, a_2, b_1, b_2)$. In this case, we add the edge $a_{max} b_1$ if $w(a_2) + w(b_2) \ge w(a_{max}) + w(b_1)$ or the edge $a_2 b_2$, otherwise (procedure \emph{KillC4}$(a_{max}, a_2, b_1, b_2)$). The vector $d[\cdot]$ is updated accordingly and the entire operation can be accomplished in time ${\cal O}(|K_i \cup K_{i+1}|)$.

\medskip\noindent
Case~\emph{(iii)} applies when one of two subcases arises. In subcase~\emph{(iii-a)} $d[a_{max}] = |B| - 1$, $b_1$ is the unique node in $B$ non-adjacent to $a_{max}$, $a_2$ is any node in $A$ adjacent to $b_1$ and, contrary to the previous case, $d[a_2] \le |B| - 2$; then we let $\bar a = a_2$. In subcase~\emph{(iii-b)} $d[a_{max}] \le |B| - 2$ and we let $\bar a = a_{max}$.

In both cases, $\bar a$ has at least two non-adjacent nodes in $B$ and we add all the edges $\bar a b$ with $b \in B \setminus N(\bar a) \cup \{b^*\}$, where $b^*$ is the node in $B \setminus N(\bar a)$ having maximum weight (procedure \emph{KillDiags}$(\bar a)$). We also update the vector $d[\cdot]$ accordingly. This task can be accomplished in ${\cal O}(|K_i \cup K_{i+1}|)$ time. Observe that in subcases~\emph{(iii-a)} and \emph{(iii-b)} the pair $(\bar A, \bar B)$ with $\bar A = \{\bar a\}$ and $\bar B = B \setminus N(\bar a)$ is a semi-homogeneous pairs of cliques as described in \emph{(i)} of Theorem~\ref{C4Properties} and hence the added edges are diagonals of squares in $G$. Consequently, by \emph{(ii)} of the same theorem, the addition of the new edges preserves claw-freeness and the property that $(A, B)$ is a square-semi-homogeneous pair of cliques.

\medskip\noindent
In each iteration of the inner loop, Case~\emph{(i)} can occur at most $|K_i|$ times. Moreover, the occurrence of Case~\emph{(ii)} makes a node in $A$ universal to $B$. It follows that also Case~\emph{(ii)} can occur at most ${\cal O}(|K_i|)$ times. Finally, the occurrence of case~\emph{(iii)} increases $d[\bar a]$ for some node $\bar a \in A$ from a value less than or equal to $|B| - 2$ to $|B| - 1$. Hence also Case~\emph{(iii)} can occur at most ${\cal O}(|K_i|)$ times.

\medskip\noindent
In conclusion, the number of stages of the inner loop is ${\cal O}(|K_i|)$. Moreover, by the above discussion we have that each stage is performed in ${\cal O}(|K_i \cup K_{i+1}|)$ time. It follows that the inner loop has a complexity of ${\cal O}((|K_i \cup K_{i+1}|)^2)$  time while the overall complexity of Procedure~\emph{Interval($G$)} is ${\cal O}(\sum_{i = 1}^{p-1}(|K_i \cup K_{i+1}|)^2) = {\cal O}(|E|))$ time.

\bigskip\noindent
\begin{procedure}
\label{Interval}
\small
\SetKwFunction{ArgMax}{ArgMax}
\SetKwFunction{Remove}{Remove}
\SetKwFunction{KillDiags}{KillDiags}
\SetKwFunction{KillCfour}{KillC4}
\SetKwFunction{d}{d}
\KwIn{A claw-free clique-strip $H$ defined by $\{K_1, \dots, K_p\}$ and a node-weighting vector $w$.}
\KwOut{A graph with no square (interval) obtained by adding new edges to $H$.}
%\Begin{
  \For {$i = 1, \dots, p-1$} {
      $A \leftarrow K_i \cap N(K_{i+1})$\;
      $B \leftarrow K_{i+1} \cap N(K_i)$\;
      \lForEach {$a \in A$} {
         $\d[a] \leftarrow |N(a) \cap B|$\;
      }
      \Repeat(Stage){$A = \emptyset$} {
         $a_{max} \leftarrow \ArgMax_{a \in A}\{\d[a]\}$\;
         \If{$\d[a_{max}] = |B|$} {
            \tcp{Case \emph{(i)}}
            $\Remove (a_{max})$\;
         }
         \ElseIf{$\d[a_{max}] = |B| - 1$} {
            $\{b_1\} \leftarrow B \setminus N(a_{max})$\;
            select $a_2 \in N(b_1) \cap A$\;
            \If{$\d[a_2] = |B| - 1$} {
               \tcp{Case \emph{(ii)}}
               $\{b_2\} \leftarrow B \setminus N(a_2)$\;
               $\KillCfour (a_{max}, b_1, a_2, b_2)$\;
            }
            \Else {
               \tcp{Case \emph{(iii-a)}}
               $\KillDiags (a_2)$\;
            }
         }
         \Else {
            \tcp{Case \emph{(iii-b)}}
            $\KillDiags (a_{max})$\;
         }
      }
   }
%}
\caption{Interval($H$)}
\end{procedure}

\medskip\noindent
In \cite{OrioloetAl08} Oriolo et al.\ prove a useful lemma concerning semi-homogeneous pairs of cliques. The following is a rephrasing of that lemma.

\begin{Lemma}\label{SemiHomogeneousLemma}
Let $(A, B)$ be a semi-homogeneous pair of cliques in a graph $G(V, E)$ and let $v$ be a node in $V \setminus (A \cup B)$. Then $v$ is either adjacent to every stable set of $G[A \cup B]$ of cardinality $2$ or to none of them.
\end{Lemma}

\noindent
\emph{Proof}. If $v$ is universal to $A$ or to $B$ then it is trivially adjacent to every stable set of $G[A \cup B]$ of cardinality $2$ and we are done. Hence assume that $v$ is neither universal to $A$ nor to $B$. Consequently, since $(A, B)$ is a semi-homogeneous pair of cliques, $v$ is null to $A \cup B$. The lemma follows.
\done

\begin{Theorem}\label{PreservingAlpha}
Let $G(V, E)$ be a node-weighted \{claw, net\}-free graph and $X$ a clique in $G$ such that $G - X$ is partitioned in two (possibly coincident) clique-strips. For each node $v \in X$, let $G^v = G - N[v]$. Let $\overline{G}$ be the interval graph obtained from $G - X$ by applying Procedure~\emph{Interval} and $\overline{G^v} = \overline{G} - N[v]$ ($v \in X$). Then $\alpha_w(G - X) = \alpha_w(\overline{G})$ and $\alpha_w(G^v) = \alpha_w(\overline{G^v})$, for each $v \in X$.
\end{Theorem}

\noindent
\emph{Proof}. The procedure \emph{Interval} produces a sequence of graphs $(G_0, \dots G_q)$ with $G_0 \equiv G - X$ and $G_q \equiv \overline{G}$. Each graph $G_i$ ($i = 1, \dots, q$) is obtained from $G_{i-1}$ in a single iteration of the inner loop of Procedure~\emph{Interval}. In particular, for some semi-homogeneous pair of cliques $(A_i, B_i)$ in $G_{i-1}$, we turn into edges all the pairs of non-adjacent nodes in $G_{i-1}[A_i \cup B_i]$ with the exception of pair $\{u, v\}$ which maximizes $w(u) + w(v)$. Evidently, $\alpha_w (G_{i-1}) \ge \alpha_w (G_i)$. Moreover, for each $v \in X$, let $G^v_i = G_i \setminus N[v]$. We have $G^v_0 \equiv G^v$ and $G^v_q \equiv \overline{G^v}$.

\smallskip\noindent
We first prove that $\alpha_w(G^v) = \alpha_w(\overline{G^v})$. Conversely, suppose $\alpha_w(G^v) > \alpha_w(\overline{G^v})$. Let $j \ge 1$ be the smallest index with the property that $\alpha_w(G^v) > \alpha_w(G^v_j)$ and let $S$ be a maximum weight stable set of $G^v_{j-1}$. Since $S$ is not a stable set of $G^v_j$, there exist nodes $a \in A_j \cap S \setminus N[v]$ and $b \in B_j \cap S \setminus N[v]$ with the property that $ab$ is a new edge in $G^v_j$. Let $\{a', b'\}$ with $a' \in A_j$, $b' \in B_j$ the pair which maximizes $w(a') + w(b')$. We have that $a'$, $b'$ are not adjacent in $G_j$ and $w(a') + w(b') \ge w(a) + w(b)$. Since $v$ is non-adjacent to $a$ and $b$, by Lemma~\ref{SemiHomogeneousLemma} $v$ is also non-adjacent to $a'$ and $b'$. It follows that $a'$ and $b'$ belong to $G^v_j$. But then, again by Lemma~\ref{SemiHomogeneousLemma}, the set $S \setminus \{a, b\} \cup \{a', b'\}$ is a stable set of $G^v_j$ having weight not smaller than $\alpha_w(G^v_{j-1})$, a contradiction.

\smallskip\noindent
The same arguments used above show, in the special case in which $N(v) \setminus X = \emptyset$, that $\alpha_w(G - X) = \alpha_w(\overline{G})$ and the theorem follows.
\done

\medskip\noindent
We conclude by giving a streamlined description of the procedure for solving the Maximum Weight Stable Set Problem in a \{claw, net\}-free graph $G(V, E)$ with $\alpha(G) \ge 4$ in time ${\cal O}(|V|\sqrt{|E|})$. First, using the results of \cite{NobiliSassano15a} we either verify that $\alpha(G) \le 3$ or construct a stable set $S_0$ of cardinality at least $4$ in time ${\cal O}(|E|)$. Next we construct in ${\cal O}(|E|)$ time a canonical stable set of cardinality at least $4$ from $S_0$. Subsequently, by Theorem~\ref{CliqueStripTheorem}, in ${\cal O}(|E|)$ time
%
%%%%%%%%%%%%%%%%%%%%%%%%%%%%%%%%%%%%%%%%%%%%%%%%%%%
%
%   Risposta al bullet 4 di Referee 1
%
we find a clique $X$ such that $G - X$ is the union of at most two clique-strips defined by clique families ${\cal H} = \{H_1, \dots, H_p\}$ and ${\cal K} = \{K_1, \dots, K_t\}$. Finally, we apply Procedure~\emph{Interval} to the clique-strips in $G - X$ and turn $G - X$ into a graph $\overline{G}$ in ${\cal O}(|E|)$ time. By Lemma~\ref{ConsistentOrdering}, $\overline{G}$ is an interval graph whose consistent ordering can be found in time ${\cal O}(|E| + |V|\log|V|)$. Now, for each node $v \in X$, let $G^v = G - N[v]$ and $\overline{G^v} = \overline{G} - N[v]$. Evidently, $\alpha_w(G) = \max \{\alpha_w(G - X), \max_{v \in X} \{\alpha_w(G^v) + w(v)\}\}$. By Theorem~\ref{PreservingAlpha}, we have $\alpha_w(G - X) = \alpha_w(\overline{G})$ and $\alpha_w(G^v) = \alpha_w(\overline{G^v})$, for each $v \in X$. Since $\overline{G^v}$ ($v \in X$) is an induced subgraph of $\overline{G}$, it is also an interval graph and inherits from $\overline{G}$ the consistent ordering of its nodes. Hence, the maximum weight stable set problem on $G$ can be reduced to solving $|X| + 1$ maximum weight stable set problems on interval graphs.

\medskip\noindent
In \cite{ManninoORC07} Mannino, Oriolo, Ricci and Chandran proved that, given a consistent ordering, the maximum weight stable set problem on an interval graph $G(V, E)$ can be solved in ${\cal O}(|V|)$ time. It follows that the maximum weight stable set problem in a connected \{claw, net\}-free graph $G(V, E)$ with $\alpha(G) \ge 4$ can be solved in time ${\cal O}(|E| + |V|\log|V| + (|X| + 1)|V|) = {\cal O}(|V|\sqrt{|E|})$.

\section*{Acknowledgement}

Our deepest thanks are due to the anonymous referees whose insightful suggestions have allowed us to reshape the paper and improve the quality of the presentation.

\section*{References}

\bibliographystyle{elsarticle-num}
\bibliography{ClawFreeNS}

\end{document}